\documentstyle[prd,eqsecnum,aps]{revtex}

\begin{document}

\title
{Supersymmetric homogeneous quantum cosmologies\\
coupled to a scalar field}
\author{J.~Bene \thanks{Permanent address: Institute for
Solid State Physics, E\"otv\"os University, Muzeum krt. 6-8,
H-1088 Budapest, Hungary} and R. Graham\\
Fachbereich Physik\\ Universit\"at Gesamthoschule Essen\\
D4300 Essen 1, Germany}
\date{~~}
\maketitle

\begin{abstract}
Recent work on $N=2$ supersymmetric Bianchi type IX cosmologies
coupled to a scalar field is extended to a general treatment of
homogeneous quantum cosmologies with explicitely solvable momentum
constraints, i.e. Bianchi types I, II, VII, VIII besides the Bianchi
type IX, and special cases, namely the Friedmann universes, the
Kantowski-Sachs space, and Taub-NUT space. Besides the earlier
explicit solution of the Wheeler DeWitt equation for Bianchi type IX,
describing a virtual wormhole fluctuation, an additional explicit
solution is given and identified with the `no-boundary state'.
\end{abstract}

\section{Introduction}\label{sec:1}

Quantum cosmology is the application of quantum mechanics to the
earliest universe. General relativity predicts that the earliest
universe is dominated by gravity, which is enhanced above all other
interactions, up to a singularity, by the strong nonlinearity (i.e.
gravitation of gravity) built into that theory. Quantum cosmology
therefore must contain a theory of quantum gravity. Unfortunately a
consistent quantum field theory of gravity has not yet been given.
However, some of the questions of quantum cosmology (but, of course,
not all of them, and possibly not the deepest ones) can already be
raised in `toy models' of quantum gravity which bypass the unsolved
problems of quantum gravitational field theory and instead require
only the framework of quantum mechanics of a system with a finite
number of degrees of freedom (see e.g. [1-7]).

Such toy models can be constructed by quantizing not the full theory
of general relativity (or even larger theories in which it is
contained as some limiting case) but only certain classes of its
spatially homogeneous solutions. The restriction to spatial
homogeneity implies, in principle, a dimensional reduction of the
(1+3)-dimensional field theory of gravitation down to a
(1+0)-dimensional quantum mechanical model. The configuration space
of the reduced models is no longer the full function space (called
superspace \cite{A}) of all 3-metrics on a space-like 3-surface
embedded in space-time \cite{B}, but merely the finite-dimensional
space (called mini-superspace \cite{C}) spanned by the parameters of
the considered class of homogeneous 3-metrics. Such reduced models
avoid the unsolved problems associated with the short-wavelength
limit of quantum gravity; on the other hand they can, of course, also
shed no light on these problems.

Spatially homogeneous 3-metrics evolving along some trajectory
through mini-superspace are special exact solutions of classical
general relativity \cite{D,E}. Whether corresponding exact solutions
of quantum gravity exist cannot be answered with confidence before
such a theory has been constructed. The dimensionally reduced quantum
mechanical mini-superspace models cannot, therefore, be considered to
be exact solutions like their classical counterparts, but must be
considered as just models. Their usefulness and interest hinges on
the fact that they provide a comparatively transparent framework in
which some questions raised by quantum cosmology can be studied (see
e.g. [1-7]).

The dimensional reduction from a field theory down to a finite number
of degrees of freedom can, in principle, be carried out before or
after quantizing the theory, and the result need not be the same. In
the present paper we shall follow the first path: As we consider
mini-superspace as a device to avoid the problems of the
gravitational quantum field theory it would seem inconsistent with
that reasoning to invoke the quantized field theory, if only
formally, in an intermediate step. Even more important, from a
practical point of view, is the fact that model-building is easier
and more direct starting from the classically reduced theory -- and,
as we have mentioned, we are not trying to construct solutions of
quantum gravity, but models.

Spatially homogeneous 3-geometries are 3-manifolds on which one of
the three-dimensional Lie groups acts transitively \cite{D,E}. Thus,
on the 3-manifold there are three linearly independent Killing
vectors $\xi_i$, $i= 1, 2, 3$, satisfying the Lie algebra
$[\xi_i,\xi_j]= {{\cal C}^k}_{ij}\xi_k$ where
${{\cal C}^k}_{ij}=-{{\cal C}^k}_{ji}$ are the structure constants of
the group. The three-dimensional Lie groups have been classified by
Bianchi \cite{H} into 9 different types (see e.g. \cite{E}),
according to their structure constants. A complete list of the
structure constants, an invariant basis $\chi_i$ and its dual basis
of one-forms $\omega^i$, $i= 1, 2, 3$, and a complete set of Killing
vectors $\xi_i$ for all Bianchi types has been given by Taub \cite{I}
(see also \cite{E}). The 4-metric of a spatially homogeneous
space-time can be written in the form
\begin{equation}
ds^2=-N^2(t) dt^2+g_{ij}(t)\omega^i\omega^j
\label{eq:1-1}
\end{equation}
where the basis one-forms $\omega^i$ satisfy
\begin{equation}
d\omega^i=\frac{1}{2}{{\cal C}^i}_{jk}\omega^j\wedge\omega^k.
\label{eq:2-1}
\end{equation}
$N(t)$ is the lapse-function which is an arbitrary positive function
of time and reflects the reparametrization invariance of the time
coordinate $t$. The elements of the tensor $g_{ij}$ of the
three-metric in the basis $\omega^i$ depend only on time and not on
the spatial coordinates.  Therefore, there are at most 6 independent
elements of $g_{ij}(t)$, spanning the space of all allowed 3-metrics,
the mini-superspace.  Here we shall restrict our attention to the
case where $g_{ij}(t)$ can be consistently chosen to be diagonal. It
can then be parametrized by three parameters $\alpha,
\beta_+,\beta_-$ or $\beta^1,\beta^2,\beta^3$ via
\begin{equation}
g_{ij}(t)=\frac{1}{6\pi}e^{2\alpha(t)}\left(e^{2\bbox{\beta}(t)}
     \right)_{ij}
    = \frac{1}{6\pi}e^{2\beta^i}\delta_{ij}
\label{eq:3-1}
\end{equation}
with the diagonal traceless matrix $\bbox{\beta}(t)$ in Misner's
parametrization \cite{C,J}, $\bbox{\beta}(t)= \text{ diag }
(\beta_++\sqrt{3}\beta_-,
\beta_+-\sqrt{3}\beta_-, -2\beta_+)$. Thus mini-superspace is spanned
by merely three coordinates. The spatially homogeneous form
(\ref{eq:1-1}) of the 4-metric can be inserted into the vacuum
equations of general relativity, $R_{\mu\nu}=0$, where $R_{\mu\nu}$
is the Ricci tensor associated with the 4-metric. The resulting
equations of motion for the independent elements of $g_{ij}(t)$ are
second order in time. For some (but not all \cite{E}) of the Bianchi
types these equations of motion may be coached into the framework of
an unconstrained Hamiltonian system \cite{C}
\begin{equation}
  \dot{q}^\nu=\frac{\partial H(q,p)}{\partial p_\nu}\quad ,\quad
     \dot{p}_\nu=-\frac{\partial H(q,p)}{\partial q^\nu}
\label{eq:4-1}
\end{equation}
moving on the surface of vanishing `energy' $H$,
\begin{equation}
  H=0 .
\label{eq:5-1}
\end{equation}
In equation (\ref{eq:4-1}) the dot denotes differentiation with
respect to a suitable parameter $\lambda$ playing the role of time.
$q^\nu$ is the chosen parametrization of the independent elements of
$g_{ij}(t)$. The $p_\nu$ are the canonically conjugate momenta.
$H(q,p)$ is the Hamiltonian which generates the required equations of
motion. Choosing $d\lambda=\sqrt{3\pi/2}e^{-3\alpha}N(t)dt=
\sqrt{3\pi/2}e^{-(\beta^1+\beta^2+\beta^3)}N(t)dt$ it takes the form
\cite{C,D,E}
\begin{eqnarray}
H&=&\frac{1}{2}(-p_\alpha^2+p_+^2+p_-^2)+V^{(0)}
    (\alpha,\beta_+,\beta_-)\nonumber\\
&=&  \frac{3}{2}\left[(p_1)^2+(p_2)^2+(p_3)^2-
     2p_1p_2-2p_1p_3-2p_2p_3\right]+ V^{(0)}(\beta^1,\beta^2,\beta^3)
\label{eq:6-1}
\end{eqnarray}
with the potential
\begin{equation}
  V^{(0)}=-12\pi^2\,{^{(3)}\!g}\;{^{(3)}\!R}
\label{eq:7-1}
\end{equation}
where ${^{(3)}\!g}$ is the determinant and ${^{(3)}\!R}$ the scalar
curvature of the 3-metric. The condition $H=0$ expresses the
reparametrization invariance of the time-parameter in eq.
(\ref{eq:4-1}).

In the cases where the unconstrained description (\ref{eq:4-1}) is
valid it can be obtained most directly \cite{K} from the Hamiltonian
formulation of general relativity \cite{L}. For later convenience we
now list the Bianchi types where the unconstrained description
(\ref{eq:4-1} ) applies, together with their potentials \cite{E}
\begin{eqnarray}
   \text{ type I: }& & {{\cal C}^i}_{jk}=0\,;\quad V_1^{(0)}
       \equiv 0\label{eq:8-1}\\
\text{ type II: } & & {\cal C}^3_{12}=-{\cal C}^3_{21}=1\,;
       \quad \text{ all other }\quad{{\cal C}^i}_{jk}
  \text{ vanish; }\nonumber\\
& &  V_2^{(0)}=\frac{1}{6}e^{4\alpha}e^{-8\beta_+}=
	 \frac{1}{6}e^{4\beta^3}\label{eq:9-1}\\
\text{ type VII: } & & {\cal C}^2_{13}=-{\cal C}^2_{31}=
   {\cal C}^1_{32}=-{\cal C}^1_{23}=1\,;\quad \text{ all other }
   \quad{{\cal C}^k}_{ij}\text{ vanish; }\nonumber\\
& &  V_7^{(0)}=\frac{1}{3}e^{4\alpha} e^{4\beta_+}
	  (\cosh(4\sqrt{3}\beta_-)-1)=
\frac{1}{6}\left(e^{2\beta^1}-e^{2\beta^2}\right)^2 \label{eq:10-1}\\
\text{ type VIII: } & & {\cal C}^3_{12}=-{\cal C}^3_{21}=
  {\cal C}^1_{32}= -{\cal C}^1_{23}={\cal C}^2_{31}=
   -{\cal C}^2_{13}=1; \nonumber\\
& & V_8^{(0)}=\frac{1}{6}e^{4\alpha}
   \Bigg[2e^{4\beta_+} (\cosh(4\sqrt{3}\beta_-)-1)
   +e^{-8\beta_+}\nonumber\\
& &\qquad\quad+4e^{-2\beta_+}\cosh(2\sqrt{3}\beta_-)\Bigg]\nonumber\\
 & & =\frac{1}{6}\Bigg[ e^{4\beta^1}+e^{4\beta^2}+e^{4\beta^3}
   +2e^{2\beta^3}(e^{2\beta1}+e^{2\beta^2})-
    2e^{2\beta^1+2\beta^2}\Bigg] \label{eq:11-1}\\
\text{ type IX: } && {\cal C}^1_{23}=-{\cal C}^1_{32}=
   {\cal C}^2_{31}= -{\cal C}^2_{13}={\cal C}^3_{12}=-
     {\cal C}^3_{21}=1;\nonumber\\
& & V_9^{(0)}=\frac{1}{6}e^{4\alpha} \Bigg[2e^{4\beta_+}
   (\cosh(4\sqrt{3}\beta_-)-1) +e^{-8\beta_+}\nonumber\\
& & \qquad\quad-4e^{-2\beta_+}\cosh(2\sqrt{3}\beta_-)\Bigg]\nonumber\\
& & = \frac{1}{6}\Bigg[ e^{4\beta^1}+e^{4\beta^2}+
e^{4\beta^3}-2e^{2\beta^1+2\beta^2}-
2e^{2\beta^1+2\beta^3}-2e^{2\beta^2+2\beta^3}\Bigg]
\label{eq:12-1}
\end{eqnarray}
The other Bianchi types and more general cases of type VII, in
general do not give rise to unconstrained Hamiltonian systems.

We add to this list a few special cases where further symmetries are
present and where an unconstrained Hamiltonian system with less than
three degrees of freedom is obtained.

\noindent
Friedmann-Robertson-Walker (FRW) universe:\\ The closed $(k=1)$, open
$(k=-1)$, and flat $(k=0)$ FRW universes classically, are isotropic
special cases of the Bianchi types IX, V (${\cal C}^1_{13}=-{\cal
C}^1_{31}={\cal C}^2_{23}= -{\cal C}^2_{32}=1$ all other ${{\cal
C}^k}_{ij}$ vanish), and I, respectively, where $p_+=p_-=0$,
$\beta_+=\beta_-=0$. Thus for the FRW universe without matter
\begin{eqnarray}
 H_{\text{ FRW }} &=& -\frac{p_\alpha^2}{2}+V_{\text{ FRW }}^{(0)}
  (\alpha)\nonumber\\
 V_{\text{ FRW }}^{(0)} &=& -\frac{k}{2}e^{4\alpha}
\label{eq:13-1}
\end{eqnarray}
It should be noted that the closed FRW universe without matter cannot
exist classically.

\noindent
Kantowski-Sachs (KS) models \cite{M}:\\ These spaces have a
4-dimensional symmetry group with a three-dimensional subgroup which
is of Bianchi type IX \cite{E}.  However, the latter subgroup does
not act transitively in three-space but only on two-dimensional
surfaces foliating the three space. The space-time metric may be
written in the standard form (\ref{eq:1-1}), (\ref{eq:3-1}) with
\begin{eqnarray*}
   \omega^3 &=& dr\;,\quad\omega^1=d\theta\;,\quad\omega^2=
       \sin\theta d\varphi\\ \bbox{\beta}(t)&=& \text{ diag }
	  (\beta_+,\beta_+,-2\beta_+)
\end{eqnarray*}
and
\begin{equation}
  \begin{array}{l} H_{\text{ KS }}=
-\frac{p_\alpha^2}{2}+\frac{p_+^2}{2}+V_{\text{ KS }}^{(0)}
   (\alpha,\beta_+)= -3p_1p_3+\frac{3}{2}(p_3)^2
    + V_{\text{ KS }}^{(0)}(\beta^1,\beta^3)\\
V_{\text{ KS }}^{(0)} = -\frac{2}{3}e^{4\alpha}
     e^{-2\beta_+}= -\frac{2}{3}e^{2(\beta^1+\beta^3)}\end{array}
\label{eq:14-1}
\end{equation}

\noindent
Taub-NUT space \cite{N,I,O,E}:\\ The Taub space \cite{I} is of
Bianchi type IX with a rotational symmetry around one spatial axis.
Therefore we may take $\beta_-=0$, $p_-=0$ and have
\begin{eqnarray}
H_{\text{ T }} &=&-\frac{p_\alpha^2}{2}+\frac{p_+^2}{2}
  +V_{\text{ T }}^{(0)} (\alpha,\beta_+)
   =-3p_1p_2+\frac{3}{2}(p_2)^2+V_{\text{ T }}^{(0)}
   (\beta^1,\beta^3) \nonumber\\
V_{\text{ T }}^{(0)} &=& \frac{1}{6} e^{4\alpha}(e^{-8\beta_+}-
  4e^{-2\beta_+}) = \frac{1}{6}\left(e^{4\beta^3}-
   4e^{2\beta^1+2\beta^3}\right)
\label{eq:15-1}
\end{eqnarray}
The new parametrization
\begin{eqnarray*}
   \gamma &=& e^{\alpha+\beta_+}\\ g &=& e^{2\alpha-4\beta_+}
\end{eqnarray*}
and the redefinition of the lapse function $N(t)\rightarrow
N(t)/g(t)$ gives rise to the new Hamiltonian
\begin{equation}
   H_{\text{ T }-\text{ NUT }}=6(gp_g^2-\gamma
p_gp_\gamma)+\frac{1}{6}(g-4\gamma^2)
\label{eq:16-1}
\end{equation}
which describes a dynamical system in which the variable $\gamma$
remains positive if it is positive initially (see e.g. \cite{E})and
the variable $g$ may pass through zero. The region $g>0$ corresponds
to Taub space \cite{I}, the region $g<0$ corresponds to NUT-space
\cite{O}, in which $\omega^3$ is time-like and $dt$ space-like. As is
well known both spaces form different parts of the single Taub-NUT
space-time \cite{N,E}.

In the present paper we shall be concerned with the construction of
specific quantized versions of the dynamical systems
(\ref{eq:4-1})-(\ref{eq:7-1}). These quantum models will be
constructed by coupling additional fermionic degrees of freedom to
the purely gravitational systems (\ref{eq:4-1})-(\ref{eq:7-1}) in
such a way that the coupled system acquires a larger symmetry, namely
supersymmetry. For this to be possible the potentials (1.8)-(1.16)
must satisfy a certain condition; they must be derivable from an
underlying (usually simpler) potential $\phi$. This condition is
verified for all the listed Bianchi types in section \ref{sec:2}.

A further ingredient in the construction of our quantum models is the
coupling of the supersymmetrized gravity model to a supersymmetric
spatially homogeneous matter field, represented by a spatially
homogeneous complex scalar field and its supersymmetric fermionic
partner.

A part of the theory we shall describe here for all the Bianchi types
listed has been presented for the Bianchi type IX in our earlier
paper \cite{P} which, in turn, builds on our earlier work in
\cite{Q,R,S}.  An application of the theory to the Bianchi type II
(but without coupling to a matter field) has already been given in
\cite{T}.

Dimensional reductions from (1+3)-dimensional supergravity down to
(1+0)-dimensional supersymmetric theories have been presented for the
Bianchi type I model without matter \cite{U,V} the closed Friedmann
model without \cite{W} and with matter \cite{X}, and the Taub model
\cite{Y}. From (1+3)-dimensional supergravity with a single
conserved real spinorial supercharge $(N=1)$ a (1+0)-dimensional
theory with $N=4$ conserved real supercharges is obtained.

By contrast, the general supersymmetric extension of the Hamiltonian
systems (\ref{eq:4-1})-(\ref{eq:7-1}) we shall consider in this paper
only leads to an $N=2$ supersymmetry. In principle, it may be
considered as a subsymmetry of the larger $N=4$ supersymmetry
obtained from supergravity, but we shall not attempt here to make
that connection explicit. The further extension of our models to
$N=4$ supersymmetry is nontrivial and requires further work. For the
case of the Bianchi type IX model without matter such an extension is
given in \cite{Z}.

\section{Supersymmetric extension of Bianchi types}\label{sec:2}
\subsection{Supersymmetry condition of the Bianchi potentials}
 \label{subsec:2a}

The geometrodynamics of the Bianchi types considered reduce,
formally, to the Hamiltonian dynamics of a particle in a
3-dimensional potential. (Eliminating the arbitrary time-parameter
the description may even be further reduced to motion in a
2-dimensional time-dependent potential, but we shall not make use of
this possibility here).  However, an important non-standard feature
in this picture is the non-definite metric $G^{(0)}_{\mu\nu}$ in
mini-superspace, whose line-element in the parametrization by
$\alpha,\beta_+,\beta_-$ or $\beta^1,\beta^2,\beta^3$ may be written
as
\begin{equation}
   dS^2=G_{\mu\nu}^{(0)}dq^\mu dq^\nu
\label{eq:1-2}
\end{equation}
with $G_{\mu\nu}^{(0)}=\text{ diag } (-1, 1, 1)$ or
$G_{\mu\nu}^{(0)}=-\frac{1}{6}(1-\delta_{\mu\nu}$), respectively. In
fact, as emphasized by Misner \cite{C}, due to reparametrization
invariance with respect to
$\lambda,d\lambda=e^{2\Omega}d\lambda^{(0)}$, the metric in
minisuperspace is fixed only up to an arbitrary conformal factor,
here written as $\exp (2\Omega(q))$,
\begin{equation}
   G_{\mu\nu}=e^{2\Omega(q)}G_{\mu\nu}^{(0)}
\label{eq:2-2}
\end{equation}
The inverse of this conformal factor appears in the potential of
(\ref{eq:6-1}) i.e.
\begin{equation}
   V(q) = e^{-2\Omega(q)}V_0(q).
\label{eq:3-2}
\end{equation}
The supersymmetric extension of particle-motion in a potential well
is treated by supersymmetric quantum mechanics invented by Nicolai
\cite{a}, Witten \cite{b} and developed further by many authors (see
e.g. [29-34]). In particular, the particle dynamics in a potential on
a curved manifold (configuration space) of arbitrary dimension with
metric
\begin{equation}
   dS^2=G_{\mu\nu}(q)dq^\mu dq^\nu
\label{eq:4-2}
\end{equation}
has been studied in the $(N=2)$-supersymmetric $\sigma$-model by a
number of authors [32-34]. Supersymmetry requires that the potential
$V(q)$ is derivable from a globally defined superpotential $\phi(q)$
via
\begin{equation}
   V(q) = \frac{1}{2}G^{\mu\nu}(q)
   \frac{\partial\phi(q)}{\partial q^\mu}
     \frac{\partial\phi(q)}{\partial q^\nu}
\label{eq:5-2}
\end{equation}
We shall demand that $\phi(q)$ solves eq. (\ref{eq:5-2}) and has the
same symmetries as $H_0$. Here $G^{\mu\nu}(q)$ is the inverse of the
metric tensor in configuration space defined by the kinetic energy
\begin{equation}
   T(q,\dot{q})=\frac{1}{2}G_{\mu\nu}(q)\dot{q}^\mu\dot{q}^\nu.
\label{eq:6-2}
\end{equation}

We note that eq. (\ref{eq:5-2}) is the Hamilton-Jacobi equation
corresponding to eqs. (\ref{eq:5-1}), (\ref{eq:6-1}) in Euclidean
time, i.e. $\phi(q)$ is a Euclidean action of classical general
relativity. Depending on the boundary conditions posed there are
different solutions of the Euclidean Hamilton-Jacobi equation. The
physical interpretation of the Euclidean actions is related to
quantum tunnelling \cite{i}: they are the actions required for a
system to reach a classically inaccesible point from a given `initial
point'.  The choice of the `initial point' depends on the physical
question posed. For tunnelling out of some equilibrium state the
`initial point' will correspond to a local or global minimum of
$\phi(q)$. E.g.  for cosmology initial points corresponding to minima
of $\phi$ at fixed $\alpha$ in the limit of vanishing scale parameter
$e^\alpha\rightarrow 0$ are of particular interest. Once $\phi(q)$ is
given, the most likely path followed in the tunnelling process is
given by the solutions of the classical Euclidean equations
\begin{equation}
 p_\nu=G_{\nu\mu}\frac{dq^\mu}{d\lambda}=
    \frac{\partial\phi}{\partial q^\nu}
\label{eq:7a-2}
\end{equation}
which must be solved under the condition that the path $q(\lambda)$
connects the chosen initial point with the given final point. For a
given $\phi$ solving eq. (\ref{eq:5-2}) the solutions of eq.
(\ref{eq:7a-2}) may be used to give a physical interpretation.

If a final point is accessible from the initial point by a
classically allowed path, $\phi$ becomes imaginary. While in eq.
(\ref{eq:5-2}) this has no immediate consequence, the
supersymmetrically extended Hamiltonian then has unusual properties
which seem to indicate that supersymmetric extensions cannot be based
on imaginary or complex $\phi$ (see eq. (\ref{eq:25a-2}) below).

The new potential $\phi(q)$ is called the superpotential and appears
like a potential in the superspace version (in the sense of
supersymmetry) of the Lagrangean of the ((1+0)-dimensional)
supersymmetric $\sigma$-model.	In order to extend the Hamiltonian
(\ref{eq:6-1}) to a supersymmetric Hamiltonian we therefore have to
solve eq. (\ref{eq:5-2}) after inserting the potentials (1.8)-(1.16)
and to identify a superpotential $\phi(q)$, in each case. It follows
from eqs.  (\ref{eq:2-2}), (\ref{eq:3-2}), and (\ref{eq:5-2}) that a
superpotential $\phi(q)$ is independent of the conformal factor. It
is therefore sufficient to examine the case $\Omega(q)=0$ where the
metric $G_{\mu\nu}=G_{\mu\nu}^{(0)}$ is flat and $V(q)$ is given by
$V^{(0)}(q)$.

We shall now solve eq. (\ref{eq:5-2}) and construct the
superpotentials for the Bianchi types of section \ref{sec:1}. In the
case of Bianchi type I, $\phi(q)$ should preserve the invariance of
$H_0$ under arbitrary shifts
$\delta\alpha,\delta\beta_+,\delta\beta_-$ or
$\delta\beta^1,\delta\beta^2,\delta\beta^3$ which requires
\begin{equation}
   \phi_1\equiv 0
\label{eq:8a-2}
\end{equation}
(a constant can always be subtracted from $\phi$).

For Bianchi type II only the invariance under independent shifts
$\delta\beta^1,\delta\beta^2$ remains, which restricts the allowed
solutions of eq. (\ref{eq:5-2}) to a function of $\beta^3$ which is
obtained as
\begin{equation}
  \phi_2=\frac{1}{6}e^{2\alpha-4\beta_+}=\frac{1}{6}e^{2\beta^3}.
\label{eq:9a-2}
\end{equation}

For Bianchi type VII the symmetry operations are
$\beta^1\leftrightarrow\beta^2$,
$\beta^3\rightarrow\beta^3+\delta\beta^3$; or $\beta_-\rightarrow
-\beta_-$, $\beta_+\rightarrow\beta_++\delta\beta_+$,
$\alpha\rightarrow\alpha-\delta\beta_+$. These are respected by the
solutions
\begin{equation}
  \phi_7 = \frac{1}{3}e^{2\alpha+2\beta_+}
    (\cosh 2\sqrt{3}\beta_--A(\alpha+\beta_+))
       = \frac{1}{6}\left(e^{2\beta^1}+e^{2\beta^2}
     -2A(\frac{1}{2}(\beta_1+\beta_2))e^{\beta^1+\beta^2}
\right)
\label{eq:10a-2}
\end{equation}
where $A$ is an arbitrary function of its argument.

For Bianchi type VIII the Hamiltonian $H_0$ is invariant merely under
$\beta^1\leftrightarrow\beta^2$. This symmetry is respected by the
solutions
\begin{equation}
\phi_8=\frac{1}{6}e^{2\alpha}\left[2e^{2\beta_+}\cosh 2\sqrt{3}
   \beta_--e^{-4\beta_+}\right] =
\frac{1}{6}\left(+e^{2\beta^1}+e^{2\beta^2}-e^{2\beta^3}\right)
\label{eq:12a-2}
\end{equation}
and
\begin{eqnarray}
  \tilde{\phi}_8 &=&\frac{1}{6}e^{2\alpha} \bigg[4e^{2\beta_+}\sinh^2
   \sqrt{3}\beta_--e^{-4\beta_+}\mp4ie^{-\beta_+}
\cosh\sqrt{3}\beta_-\bigg]\nonumber\\
&=& \frac{1}{6}\bigg(e^{2\beta^1}+e^{2\beta^2}-e^{2\beta^3}-
    2e^{\beta^1+\beta^2}
\mp 2i(e^{\beta^1}+e^{\beta^2})e^{\beta^3}\bigg)
\label{eq:13a-2}
\end{eqnarray}
However, the latter solution is complex indicating that the
underlying trajectories are, in part, classically allowed. This
corresponds to the fact that $V_8^{(0)}$ is not a binding potential.

Finally, for  Bianchi type IX we have the three symmetries
$\beta^i\leftrightarrow\beta^j (i\neq j)$ . They are preserved by
\begin{eqnarray}
  \phi_9 &=& \frac{1}{6}e^{2\alpha}\left[2e^{2\beta_+} \cosh
2\sqrt{3}\beta_-+e^{-4\beta_+}\right]\nonumber\\
  &=& \frac{1}{6}\left(e^{2\beta^1}+e^{2\beta^2}+e^{2\beta^3}\right).
\label{eq:14a-2}
\end{eqnarray}
This is the superpotential chosen in ref. \cite{Q,S,Z}. As a solution
of the Euclidean Hamilton Jacobi equation it was obtained in ref.
\cite{j}. As was also shown there a further solution with the
required symmetry exists , which is given by
\begin{eqnarray}
   \tilde{\phi}_9 &=& V_9^{(0)}
\left(\frac{\alpha}{2},\frac{\beta_+}{2},\frac{\beta_-}{2}\right)
  \nonumber\\
&=& \frac{1}{6}
\left(e^{2\beta^1}+e^{2\beta^2}+e^{2\beta^3}-2e^{\beta^1+\beta^2}-
2e^{\beta^1+\beta^3}-2e^{\beta^2+\beta^3}\right).
\label{eq:15a-2}
\end{eqnarray}
Further solutions have been given in \cite{j} but they break the
permutation symmetry $\beta^i\leftrightarrow\beta^j$ and are
therefore not considered here.

\subsection{Supersymmetry condition for important special cases}
\label{subsec:2b}
Now we determine Euclidean actions for the special cases
(\ref{eq:13-1})-(\ref{eq:16-1}) listed in section \ref{sec:1}.
\noindent
FRW-Universes:\\
\begin{equation}
   G_{\mu\nu}^{(0)} = -1\;;\quad \phi_{\text{ FRW }}=
\left\{\begin{array}{lll} \frac{1}{2} & e^{2\alpha} & \quad k=1\\ 0
&	      & \quad k=0\\ \frac{i}{2} & e^{2\alpha} & \quad k=-1
\end{array}\right.
\label{eq:12-2}
\end{equation}
As the open FRW-Universe expands classically even if it is empty
$\phi$ becomes imaginary in this case. \\
\noindent
Kantowski Sachs space:\\ The Hamiltonian $H_0$ is invariant under
$\beta^1\rightarrow\beta^1+\delta\beta$,
$\beta^3\rightarrow\beta^3-\delta\beta$ i.e. $\phi$ can only depend
on $\beta^1+\beta^3$. Thus, in the variables $\alpha, \beta_+$
\begin{equation}
   G_{\mu\nu}^{(0)} = {-1~0~\choose\phantom{-}0~1~}; \quad
   \phi_{\text{ KS }}=\frac{1}{3}e^{2\alpha-\beta_+}
\label{eq:13-2}
\end{equation}
Taub space:\\ We only consider solutions obtained by restrictions of
the Bianchi type IX solutions to axial symmetry. In the variables
$\alpha,\beta_+$
\begin{equation}
   G_{\mu\nu}^{(0)} = {-1~0~\choose\phantom{-}0~1~}; \quad
    \phi_{\text{ T }}=\frac{1}{6}e^{2\alpha} \left(2e^{2\beta_+}+
      e^{-4\beta_+}\right)
\label{eq:14-2}
\end{equation}
and
\begin{equation}
  \tilde{\phi}_{\text{ T }}=\frac{1}{6}e^{2\alpha}(e^{-4\beta_+}-
    4e^{-\beta_+})
\label{eq:19a-2}
\end{equation}
Taub-NUT space:\\
\begin{equation}
   G^{\mu\nu} = {12 g~-6\gamma\choose-6\gamma~ 0~\phantom{\gamma}~};
\quad \phi_{\text{ T }-\text{ NUT }}=\frac{1}{6}(2\gamma^2+g)
\label{eq:15-2}
\end{equation}
and
\begin{equation}
  \tilde{\phi}_{\text{T}-\text{ NUT }}=\frac{1}{6}(g-4\gamma\sqrt{g})
\label{eq:20a-2}
\end{equation}
As $\phi$ and $\tilde{\phi}$ are invariant against conformal changes
of the metric in mini-superspace the expression for Taub-NUT space
follow from those for Taub space by a direct substitution of the
coordinate change after eq.  (\ref{eq:15-1}).

\subsection{Supersymmetric quantization}
\label{subsec:2c}
Having established the supersymmetry condition (\ref{eq:5-2}) in all
cases we wish to consider, we are now in a position to quantize all
models in a way which renders them $(N=2)$-supersymmetric. We simply
apply to this purpose the quantization rules of the supersymmetric
$\sigma$-model.  These quantization rules may be stated as follows
\cite{f,g}:

A classical Hamiltonian system with Hamiltonian

\begin{equation}
   H_0=\frac{1}{2}G^{\mu\nu}(q)\left(p_\nu p_\mu+
   \frac{\partial\phi}{\partial q^\mu}
    \frac{\partial\phi}{\partial q^\nu}\right)
\label{eq:16-2}
\end{equation}
is quantized by associating with $H_0$ a quantum Hamiltonian $H$,
reducing to $H_0$ in the classical limit $\hbar\rightarrow 0$, of the
form
\begin{equation}
   2H=\tilde{Q}Q+Q\tilde{Q}
\label{eq:17-2}
\end{equation}
where $Q$, $\tilde{Q}$ are linear operators satisfying
\begin{equation}
   Q^2=0=\tilde{Q}^2.
\label{eq:18-2}
\end{equation}
If the matrix $G_{\mu\nu}(q)$ is positive definite, i.e. the metric
$G_{\mu\nu}$ Riemannian, $Q$ and $\tilde{Q}$ are mutually adjoint. If
the metric $G_{\mu\nu}$ is pseudo-Riemannian (as it is in
mini-superspace) then $Q$ and $\tilde{Q}$ cannot be mutually adjoint,
but become so, if a suitable Wick rotation is performed rendering
$G_{\mu\nu}$ Riemannian but keeping $\phi$ fixed. The operators $Q$
and $\tilde{Q}$ have the explicit form
\begin{eqnarray}
   Q &=& \psi^a {e_a}^\nu(q)
\left(\pi_\nu+i\frac{\partial\phi}{\partial q^\nu}\right)\nonumber\\
\tilde{Q} &=& \bar{\psi}_a e^{a\nu}(q)
\left(\pi_\nu-i\frac{\partial\phi}{\partial q^\nu}\right).
\label{eq:19-2}
\end{eqnarray}
Here the following new quantities have been introduced:

${e_a}^\nu(q)$ is the vielbein associated with $G^{\nu\mu}(q)$ and
satisfies
\begin{equation}
   {e_a}^\nu(q){e_b}^\mu(q)\eta^{ab}=G^{\nu\mu}(q)
\label{eq:20-2}
\end{equation}
where $\eta^{ab}$ is the unit tensor, if $G_{\mu\nu}$ is Riemannian,
and the Minkowski tensor $\eta^{ab}=\text{ diag }(-1, 1,\dots,
1)$ if one eigenvalue of $G_{\mu\nu}$ is negative. Latin indices are
raised and lowered by the use of $\eta^{ab }$ and $\eta_{ab}$. The
$\psi^a$ and their adjoint $\bar{\psi}_a$ are fermionic operators
satisfying
\begin{eqnarray}
   \left[\psi^a,\psi^b\right]_+ &=& 0 = \left[
\bar{\psi}_a,\bar{\psi}_b\right]_+\nonumber\\
\left[\psi^a,\bar{\psi}_b\right]_+ &=&\delta^a_b.
\label{eq:21-2}
\end{eqnarray}
The $\pi_\nu$ are operators
\begin{equation}
\pi_\nu = -i\hbar\frac{\partial}{\partial q^\nu}+i\hbar
	   {{\omega_\nu}^b}_c\bar{\psi}_b \psi^c
\label{eq:22-2}
\end{equation}
where the spin-connections ${{\omega_\nu}^b}_c$ are functions of the
$q^\mu$ and defined by
\begin{equation}
{{\omega_\nu}^b}_c = -{e^b}_{\mu}{e_c}^\mu_{;\nu}=
  -{\omega_{\nu c}}^b.
\label{eq:23-2}
\end{equation}
Here ${e_c}^\mu_{;\nu}$ denotes the Riemann-covariant derivative of
the vielbein fields. The ${{\omega_\nu}^b}_c$ vanish identically if
the $G^{\mu\nu}(q)$ are independent of the $q^\lambda$. If the metric
$G_{\mu\nu}(q)$ is conformal to a constant metric
$G_{\mu\nu}=e^{2\Omega(q)}G_{\mu\nu}^{(0)}$ as in eq.  (\ref{eq:2-2})
with the parametrization $(\alpha,\beta_+,\beta_-)$ then

\begin{equation}
{{\omega_\nu}^b}_c=\frac{\partial\Omega}{\partial
q^\lambda}\left(e^{b\lambda}e_{c\nu}- {e_c}^\lambda{e^b}_\nu\right).
\label{eq:24-2}
\end{equation}
We note that the special operator ordering in eq. (\ref{eq:19-2})
with (\ref{eq:22-2}) is crucial to ensure that $Q$ and $\tilde{Q}$
are mutually adjoint (with respect to the invariant measure
$\sqrt{\det(G_{\mu\nu})}d^nq)$ if $G_{\mu\nu}$ is Riemannian
\cite{f,g}.

The explicit form of the Hamiltonian (\ref{eq:17-2}) follows from
(\ref{eq:19-2}). In the special case where the metric (\ref{eq:20-2})
is flat and constant it takes the form
\begin{equation}
  \begin{array}{l} H=-\frac{\hbar^2}{2}G^{\nu\mu}
\frac{\partial}{\partial q^\nu}\frac{\partial}{\partial q^\mu}+
\frac{1}{2}G^{\nu\mu} \frac{\partial\phi}
{\partial q^\nu}\frac{\partial\phi}{\partial q^\mu}\\
\hspace{1cm} +\frac{1}{2}{e_a}^\nu{e_b}^\mu
	      \frac{\partial^2\phi}{\partial q^\nu\partial q^\mu}
[\bar{\psi}^a,\psi^b]\end{array}.
\label{eq:25a-2}
\end{equation}

The last term in eq. (\ref{eq:25a-2}) makes it difficult to
accommodate imaginary or complex $\phi$ which appear in the Bianchi
type VIII case (\ref{eq:13a-2}), the open FRW case (\ref{eq:12-2})
and the NUT case (\ref{eq:20a-2}) where $g<0$. We shall therefore not
consider these cases here further.

The Hamiltonian (\ref{eq:17-2}) commutes with $Q$ and $\tilde{Q}$
\begin{equation}
   [H,Q]=0=[H,\tilde{Q}]
\label{eq:25-2}
\end{equation}
i.e. the theory is invariant under the supersymmetry transformation
\begin{equation}
   \Omega\rightarrow\Omega+[\Omega,\tilde{\epsilon}Q]+
[\tilde{Q}\epsilon,\Omega]
\label{eq:26-2}
\end{equation}
where $\epsilon$ and $\tilde{\epsilon}$ are arbitrary parameters,
anticommuting among themselves and with all fermionic variables and
commuting with bosonic variables.

\subsection{Application to cosmological models}
\label{subsec:2d}
The Schr\"odinger equation is called the Wheeler DeWitt equation
[1-3] in the present case and is the quantum analog of
(\ref{eq:5-1}). It reads
\begin{equation}
   H|\psi\rangle=0.
\label{eq:27-2}
\end{equation}
Like eq.~(\ref{eq:5-1}) it expresses the local reparametrization
invariance of the arbitrary time-parameter, which does not appear in
(\ref{eq:27-2}). Supersymmetry is a local symmetry in supergravity,
i.e.  invariance under the transformation (\ref{eq:26-2}) must be
required for arbitrary {\bf time-dependent} $\tilde{\epsilon}(t)$ and
$\epsilon(t)$ \cite{k,R}.  This imposes the constraints $Q=0$,
$\tilde{Q}=0$ on the state-vector $|\psi\rangle$
\begin{equation}
   Q|\psi\rangle=0=\tilde{Q}|\psi\rangle.
\label{eq:28-2}
\end{equation}
These constraints imply the Wheeler DeWitt equation (\ref{eq:27-2}),
but they are not equivalent to it as $Q$ and $\tilde{Q}$ are not
mutually adjoint.

The fermion number
\begin{equation}
   F=\bar{\psi}_a\psi^a
\label{eq:29-2}
\end{equation}
is conserved by $H$, $[H, F]=0$, and $[Q, F]=Q$, $[\tilde{Q},
F]=-\tilde{Q}$.  Therefore the sectors with fixed fermion numbers
$F=f (0\le f\le n$, where $n$ is the dimension of mini-superspace)
can be considered separately. We define the fermion vacuum by
\begin{equation}
   \psi^a|0\rangle = 0 \quad \mbox{for all}\quad a.
\label{eq:30-2}
\end{equation}
Then, a general state with $F=f$ takes the form
\begin{equation}
|\psi_f\rangle=\frac{1}{f!}f_{\mu_1\dots\mu_f}(q)\bar{\psi}^{\mu_1}
   \dots\bar{\psi}^{\mu_f}|0\rangle
\label{eq:31-2}
\end{equation}
with completely anti-symmetric $f_{\mu_1\dots \mu_f}$, where
\begin{equation}
   \bar{\psi}^\mu=e^{a\mu}\bar{\psi}_a.
\label{eq:32-2}
\end{equation}
In this representation $\pi_\mu$ is proportional to the covariant
derivative operator $\pi_\mu=-i\nabla_\mu$
\begin{equation}
\pi_\mu|\psi_f\rangle=\frac{1}{f!}\bar{\psi}^{\mu_1}\dots
  \bar{\psi}^{\mu_f}|0\rangle(-i)f_{\mu_1\dots\mu_f;\mu}.
\label{eq:33-2}
\end{equation}
Thus
\begin{equation}
   \begin{array}{l}
Q|\psi_f\rangle=\frac{1}{(f-1)!}\bar{\psi}^{\mu_1} \dots
\bar{\psi}^{\mu_{f-1}}|0\rangle\left(-i\hbar\nabla_\mu+i
\frac{\partial\phi}{\partial
q^\mu}\right){f^\mu}_{\mu_1\dots\mu_{f-1}}\\
\tilde{Q}|\psi_f\rangle=\frac{1}{(f+1)!}\bar{\psi}^{\mu_1} \dots
\bar{\psi}^{\mu_{f+1}}|0\rangle
\Bigg[\bigg(-i\hbar\frac{\partial}{\partial q^{\mu_1}}
 -i\frac{\partial\phi}{\partial
q^{\mu_1}}\bigg)f_{{\mu_2}\dots\mu_{f+1}}\\ \hspace{6cm}
+\;\mbox{cyclic permutations}\;\Bigg]\end{array}
\label{eq:34-2}
\end{equation}
The constraints (\ref{eq:28-2}) therefore reduce to
\begin{equation}
 \begin{array}{l} \left(-i\hbar\nabla_\mu+i
\frac{\partial\phi}{\partial q^\mu}\right)f^\mu_{\mu_1\dots
  \mu_{f-1}}(q) = 0\\
\epsilon^{\mu_1\dots\mu_{f+1}}\left(-i\hbar
  \frac{\partial}{\partial q^{\mu_1}} -i
\frac{\partial\phi}{\partial q^{\mu_1}}\right)f_{{\mu_2}\dots
  \mu_{f+1}}=0
\end{array}
\label{eq:35-2}
\end{equation}
in the $f$-fermion sector.

A similar representation can be built on the filled fermion state
$|n\rangle$ defined by
\begin{equation}
   \bar{\psi}^a|n\rangle=0\quad\mbox{for all}\quad a.
\label{eq:36-2}
\end{equation}
Thus
\begin{eqnarray}
 |n\rangle & = &\frac{1}{n!}\epsilon_{a_1\dots a_n}
\bar{\psi}^{a_1}\dots\bar{\psi}^{a_n}|0\rangle\nonumber\\
& = & \frac{1}{n!}\sqrt{|\det(G_{\mu\nu})|}
\epsilon_{\mu_1\dots\mu_n}\bar{\psi}^{\mu_1}
  \dots\bar{\psi}^{\mu_n}|0\rangle
\label{eq:37-2}
\end{eqnarray}
In this representation the state $|\psi_{n-f}\rangle$ takes the form
\begin{equation}
   |\psi_{n-f}\rangle = \frac{1}{f!}g_{\mu_1\dots\mu_f}(q)
\psi^{\mu_1}\dots\psi^{\mu_f}|n\rangle
\label{eq:38-2}
\end{equation}
The constraints (\ref{eq:28-2}) now take the form
\begin{equation}
   \begin{array}{l}
\epsilon^{\mu_1\dots\mu_{f+1}}\left(-i\hbar\frac{\partial}{\partial
q^{\mu_1}} +i\frac{\partial\phi}{\partial
q^{\mu_1}}\right)g_{\mu_2\dots\mu_{f+1}}=0\\
\left(-i\hbar\nabla_\mu-i\frac{\partial\phi}{\partial q^\mu}\right)
g^\mu_{\mu_1\dots\mu_{f-1}}=0 \end{array}
\label{eq:39-2}
\end{equation}
i.e. they have the same form as eqs. (\ref{eq:35-2}) with
$|\psi_f\rangle\Leftrightarrow|\psi_{n-f}\rangle $,
$f_{\mu_1\dots\mu_f}\Leftrightarrow g_{\mu_1\dots\mu_f}$ and
$\phi\Leftrightarrow-\phi$ interchanged.

Comparing with equation (\ref{eq:35-2}) we find, possibly up to an
irrelevant sign,
\begin{equation}
   f_{\mu_1\dots\mu_f}(q) = \sqrt{|det G_{\mu\nu}|}
\epsilon_{\mu_1\dots\mu_f\mu_{f+1}\dots\mu_n} g^{\mu_{f+1}\dots
\mu_n}.
\label{eq:40-2}
\end{equation}
Thus if the solutions in the sectors $f=0,\dots, [n/2]$ have been
determined those in the other sectors can be inferred.

If a Wick-rotation is performed in the Hamiltonian to render the
metric in mini-superspace Riemannian but keeping $\phi$ fixed, then
the classical Hamiltonian $H_0$, eq.~(\ref{eq:6-1}), becomes positive
definite. Furthermore, then
$\eta^{\nu\mu}=\text{ diag } (1, 1,\dots 1)$
holds and ${e_a}^\nu = e^{a\nu}$. Thus $Q$ and $\tilde{Q}$ are then
mutually adjoint and $H$ is self-adjoint. In this case a nontrivial
state $|\psi_f\rangle$ satisfying $\tilde{Q}|\psi_f\rangle=0$,
$Q|\psi_f\rangle=0$ cannot be written as $|\psi_f\rangle =
\tilde{Q}|\psi_{f-1}\rangle$ with another state $|\psi_{f-1}\rangle$
(which would necessarily be in the $F=f-1$ sector), or as
$|\psi_f\rangle=Q|\psi_{f+1}\rangle$ with another state
$|\psi_{f+1}\rangle$ (which would necessarily be in the $F=f+1$
sector), because otherwise
\begin{equation}
   \langle\psi_f|\psi_f\rangle = \langle\psi_{f-1}|Q|\psi_f\rangle=0
\label{eq:41-2}
\end{equation}
or
\begin{equation}
   \langle\psi_f|\psi_f\rangle =
\langle\psi_{f+1}|\tilde{Q}|\psi_f\rangle=0
\label{eq:42-2}
\end{equation}
would follow (where now the scalar product also includes an
integration over the $q$ with their invariant measure in
mini-superspace).

On the other hand, if $G_{\mu\nu}$ is pseudo-Riemannian these
conclusions do not hold because then neither $(Q)^+|\psi\rangle$ nor
$(\tilde{Q})^+|\psi\rangle$ need to vanish. A Wick-rotation in
mini-superspace therefore severely restricts the possible solutions
in supersymmetric quantum cosmology.

Formal solutions in the sectors $f=0$ and $f=n$ can be immediately
written down:
\begin{eqnarray}
   |\psi_0\rangle &=& \text{ const }\exp
(-\phi/\hbar)|0\rangle\nonumber\\ |\psi_n\rangle &=&
 \text{ const }\exp(+\phi/\hbar|n\rangle.
\label{eq:43-2}
\end{eqnarray}
These solutions are the only ones in these sectors. They apply to all
the cosmological models we have described in the previous sections
and exist independent of whether a Wick rotation has been performed
or not. However, they are formal, because no boundary conditions have
been posed so far and no criteria have been given allowing to decide
whether these solutions are physically acceptable or have to be
rejected.  Such criteria could only be derived from a definite
physical interpretation of the wave-function. Unfortunately this
point is far from being well understood and shall be discussed
further below.	However, for the time being we shall demand that a
physically acceptable state (\ref{eq:43-2}) is normalizable for fixed
$\alpha$ (scale parameter)
\cite{Q}.

As $G_{\mu\nu}$ is pseudo-Riemannian (no Wick rotation), it is
possible to construct solutions in all other fermion sectors as well
\cite{S}.  As in all examples so far we have $n\le 3$, it is enough to
consider the sector with $f=1$. With the general ansatz
\begin{equation}
   |\psi_1\rangle=f_\nu(q)e^{-\phi(q)/\hbar}\bar{\psi}^\nu|0\rangle
\label{eq:44-2}
\end{equation}
we obtain the system of equations
\begin{eqnarray}
   \hbar {f^\nu}_{;\nu}-2\phi_{;\nu} f^\nu &=& 0\nonumber\\
  \frac{\partial f_\nu}{\partial q^\mu}-
  \frac{\partial f_\mu}{\partial q^\nu} &=& 0.
\label{eq:45-2}
\end{eqnarray}
Due to the second equation there is a function $f(q)$ by which
$f_\nu$ can be represented as $f_\nu=-i\hbar\frac{\partial
f}{\partial q^\nu}$. Then the ansatz (\ref{eq:44-2}) takes the form
\begin{equation}
   |\psi_1\rangle=\tilde{Q} f(q)e^{-\phi(q)/\hbar}|0\rangle
\label{eq:46-2}
\end{equation}
which automatically satisfies $\tilde{Q}|\psi_1\rangle=0$. As we have
already mentioned, this form of $|\psi_1\rangle$ is not permitted if
$G_{\nu\mu}$ has been Wick-rotated to a Riemannian form. It is
permitted, however, for the pseudo-Riemannian mini-superspace metric.
If $f(q)\equiv\text{ const }$, $|\psi_1\rangle$ vanishes identically,
i.e. this case has to be excluded.

The remaining condition $Q|\psi_1\rangle=0$ now leads to the
condition
\begin{equation}
   \left(\hbar\nabla_\mu-2\frac{\partial\phi}{\partial
q^\mu}\right)G^{\mu\nu} \frac{\partial}{\partial q^\nu}f(q)=0
\label{eq:47-2}
\end{equation}
which is a version of the Wheeler DeWitt equation. All its solutions
$f(q)\not\equiv\text{ const }$ give rise to states in the 1-fermion
sector.  The ansatz (\ref{eq:46-2}) automatically solves one of the
supersymmetry constraints, but it also sacrifices the first-order
form of the resulting wave-equation. One advantage, however, remains:
there are no ambiguities about the appropriate operator ordering in
the second-order wave equation. Such ambiguities have already been
resolved in the explicit expressions for $Q$ and $\tilde{Q}$.

\subsection{Physical meaning of the superpotentials}
\label{subsec:2e}
In the preceding section we have seen that simple solutions
(\ref{eq:43-2}) of the supersymmetry constraints (\ref{eq:28-2})
exist in the empty and filled fermion sectors. The semi-classical
form of these solutions offers the possibility to discuss the meaning
of the two different choices of the superpotential $\phi$ and
$\tilde{\phi}$ found in sections \ref{subsec:2b} and \ref{subsec:2c}.
We shall give this discussion first for the case of Bianchi type IX,
where the solutions (\ref{eq:43-2}) are
\begin{eqnarray}
  |\psi_0\rangle &=& \text{ const }\exp(-\phi_9/\hbar)|0\rangle
   \nonumber\\
   |\tilde{\psi}_0\rangle &=& \text{ const }\exp(-\tilde{\phi}_9/\hbar)
  |0\rangle
\label{eq:48a-2}
\end{eqnarray}
and
\begin{eqnarray}
   |\psi_3\rangle &=& \text{ const }\exp(+\phi_9/\hbar)|3\rangle
     \nonumber\\
  |\tilde{\psi}_3\rangle &=& \text{ const }\exp(+\tilde{\phi}_9/\hbar)
     |3\rangle
\label{eq:49a-2}
\end{eqnarray}
The states $|\psi_3\rangle$ and $|\tilde{\psi}_3\rangle$ both diverge
for $|\beta_\pm|\to\infty$ at fixed $\alpha$ and are therefore ruled
out as acceptable physical solutions. The two remaining solutions
(\ref{eq:48a-2}) can be interpreted semi-classically: $\phi_9$ and
$\tilde{\phi}_9$ are two different, classical extremal Euclidean
actions of two different regular Riemannian space-times, both having
the 3-geometry $g_{ij}$ of eq. (\ref{eq:1-1}) with given parameters
$\alpha,\beta_+,\beta_-$ as a boundary. As discussed in
section~\ref{subsec:2a} it is possible to reconstruct these
Riemannian space-times from their extremal action, using the canonical
relations
\begin{eqnarray}
 p_\alpha &=&
   -\frac{d\alpha}{d\lambda}=\frac{\partial\phi}{\partial\alpha}
      \nonumber\\
p_+ &=&
  \frac{d\beta_+}{d\lambda}=\frac{\partial\phi}{\partial\beta_+}
   \nonumber\\
p_-&=& \frac{d\beta_-}{d\lambda}=\frac{\partial\phi}{\partial\beta_-}.
\label{eq:50a-2}
\end{eqnarray}
For the choice $\phi=\phi_9$ these equations can be integrated in
order to determine $\alpha,\beta_+,\beta_-, N$ as a function of a
suitable affine parameter (denoted by $\rho$ below) and 3 constants
of integration \cite{l}. The result is the Bianchi type IX space-time
\cite{l}
\begin{equation}
  ds^2=\frac{d\rho^2}{\sqrt{F(\rho)}}+\frac{1}{4}\rho^6\sqrt{F(\rho)}
   \Bigg[\frac{\omega^1\omega^1}{\rho^4-a_1^4}+
    \frac{\omega^2\omega^2}{\rho^4-a_2^4}+
      \frac{\omega^3\omega^3}{\rho^4-a_3^4}\Bigg]
\label{eq:51a-2}
\end{equation}
with
\begin{equation}
  F(\rho) = \left(1-\frac{a_1^4}{\rho^4}\right)
    \left(1-\frac{a_2^4}{\rho^4}\right)
     \left(1-\frac{a_3^4}{\rho^4}\right).
\label{eq:52a-2}
\end{equation}
Here $a_1, a_2, a_3$ are constants of integration. They have to be
chosen in such a way that the given 3-metric of the Bianchi type IX
space-time with parameters $\alpha,\beta_+,\beta_-$ becomes the
3-metric of eq. (\ref{eq:51a-2}) for a suitable choice of the
parameter $\rho$, with $\rho^2>\max(a_1^2, a_2^2, a_3^2)$. The
4-metric (\ref{eq:51a-2}) is regular for $\rho^2>\max(a_1^2, a_2^2,
a_3^2)$ and becomes flat Euclidean for $|\rho|\to\infty$. Therefore,
the given 3-metric must form the {\bf inner} boundary of the 4-metric
(\ref{eq:51a-2}). These boundary conditions correspond to a wormhole
\cite{m,n} with the given 3-metric arising by a quantum fluctuation
from an asymptotically Euclidean metric. The associated wave-function
vanishes rapidly in the same limit of 3-geometries (wormholes), with
large volume $(\alpha\to\infty)$, i.e. the probability amplitudes for
paths of all 3-geometries  larger than and collapsing to the given
one, whose sum builds up the wave-function $\sim\exp(-\phi_9/\hbar)$
in a path integral, interfere destructively for large 3-geometries
but constructively for small 3-geometries (on the Planck scale).
Because of the singularity of the 4-metric (\ref{eq:51a-2}) for small
$\rho^2(\rho^2=\max (a_1^2,a_2^2,a_3^2))$ it is not possible to
choose the given 3 geometry with $\alpha,\beta_+,\beta_-$ as the
outer boundary of (\ref{eq:51a-2}). Therefore the state
$\sim\exp(-\phi_9/\hbar)$ cannot describe a cosmological quantum
state arising by a sum over probability amplitudes of paths of
3-geometries {\bf expanding} into the given one.

Now we turn to the state $\sim\exp(-\tilde{\phi}_9/\hbar)$, where we
have to choose $\phi=\tilde{\phi}_9$ in eq. (2.56). The
integration of these equations now gives rise to Riemannian
space-times with 3-metrics which are regular for small 3-geometries
and the given 3-geometry must be imposed as the {\bf outer} boundary
of these space-times. This corresponds to a cosmological solution, a
quantum fluctuation `from nothing'. In fact, in a construction via a
path integral the wave-function $\exp(-\tilde{\phi}_9/\hbar)$ may be
seen as the result of a sum over the probability amplitudes of all
paths of compact 3-geometries evolving from a point to the given
3-geometry. This is precisely the prescription given by Hartle and
Hawking \cite{F} for the no-boundary state.Thus we arrive at the
remarkable conclusion that the no-boundary state for the Bianchi IX
model has an explicit and very simple form. (The state (2.54) was
first given in [50] and [17]).

Let us now discuss the solutions for the other Bianchi types in a
similar manner.

For type II eqs. (2.56) become
\begin{equation}
  \frac{d\alpha}{d\lambda}=-2\phi_2,\,
   \frac{d\beta_+}{d\lambda}=-4\phi_2,\, \beta_-=\text{ const }.
\label{eq:59a-2}
\end{equation}
and $\lambda$ is related to Euclidean standard time by
\begin{equation}
  \frac{d\lambda}{dt}=\sqrt{\frac{3\pi}{2}}e^{-3\alpha}.
\label{eq:60a-2}
\end{equation}
The solutions yield the Riemannian metrics
\begin{equation}
  ds^2=dt^2+t^2(c_1^2(\omega^1)^2+c_2^2(\omega^2)^2+
   \frac{c_3^2}{t^4}(\omega^3)^2)
\label{eq:61a-2}
\end{equation}
which describe the collapse from a disc-shaped to a pencil-shaped and
the expansion back to a disc-shaped 3-geometry from $t=-\infty$ to
$t=0$ to $t=+\infty$. The quantum fluctuation of {\bf cosmological}
interest escribed by this solution is therefore the expansion of a
pencil-shaped 3-geometry with vanishing 3-volume at $t=0$ to a given
Bianchi II 3-metric forming the outer boundary.

For type VII eqs. (2.56) are
\begin{eqnarray}
\frac{d\alpha}{d\lambda}&=&-\frac{d\beta_+}{d\lambda}=
  -2\phi_7-\frac{1}{3}e^{2\alpha+2\beta_+}A'(\alpha+\beta_+)
  \nonumber\\
\frac{d\beta_-}{d\lambda} &=&\frac{2}{\sqrt{3}}e^{2\alpha+2\beta_+}
 \sinh 2\sqrt{3}\beta_-
\label{eq:62a-2}
\end{eqnarray}
{}From the first equations it follows that $\alpha+\beta_+=$ const. The
last equation yields
\[
\beta_-=\frac{1}{\sqrt{3}}\text{ artanh }
 [{\cal C}e^{4\lambda e^{2(\alpha+\beta_+)}}]
\]
where ${\cal C}$ is a constant of integration. Therefore the value of
$\alpha+\beta_+$ is determined by the initial condition and any given
value of $\beta_-$ can be reached after making an appropriate choice
of ${\cal C}$. However, the trajectories of $\alpha$ and $\beta_+$
depend on the unspecified function $A(\alpha+\beta_+)$, i.e. the
general solution still permits a large variety of different behavior.

$\phi_8$ could be discussed similarly. However, it does not seem
possible to construct a wave-function of the form $\exp(\mp\phi_8)$
which is normalizable for fixed $\alpha$. Therefore we do not enter
that discussion here. In fact, of all the Bianchi types discussed
only the Bianchi type IX appears to give results which are of real
physical interest.

Turning to the special cases, we note first that for the closed FRW
model eq. (2.56) reduces to
\begin{equation}
  \frac{d\alpha}{d\lambda}=e^{2\alpha}\,,\,\frac{d\lambda}{dt}=
   \sqrt{\frac{3\pi}{2}}e^{-3\alpha}
\label{eq:63a-3}
\end{equation}
which gives the metric
\begin{equation}
  ds^2=dt^2+\frac{1}{4}t^2((\omega^1)^2+(\omega^2)^2+(\omega^3)^2)
\label{eq:64a-2}
\end{equation}
describing a tunnelling solution from vanishing scale parameter (at
$t=0$) to a given final value of the scale parameter. It also
describes the fluctuation from an asymptotically Euclidean 3-metric
at $t=-\infty$ to the final value of the scale parameter. The first
case may lead to the quantum initiation of an isotropic universe. The
second case may again be interpreted as a virtual wormhole.

The Taub case may be discussed as the special case of the Bianchi
type IX solutions for $\beta_-=0$. Thus $a_1=a_2$ in eqs.
(\ref{eq:51a-2}), (\ref{eq:52a-2}). It is then apparent that the
(0,0) and the (3,3) components of the Euclidean metric both change
sign when $\rho^4$ drops below $a^4$. This signals the transition
from the Taub regime to the NUT regime. This transition occurs in the
regime where eq.  (\ref{eq:51a-2}) is applicable only if
$a_1=a_2>a_3$. In the NUT-regime the analytic continuation from the
pseudo-Riemannian to a Riemannian space-time metric has to be
changed, because the $t$-coordinate becomes space-like and is not
rotated in the complex plane, while the $\omega^3$-direction becomes
time-like and must be analytically continued to the Riemannian
regime.  Thus, eq.~(\ref{eq:51a-2}) in the Taub-NUT regime becomes
\begin{equation}
ds^2=
\frac{d\rho^2}{|1-\frac{a_1^4}{\rho^4}|\sqrt{1-\frac{a_3^4}{\rho^4}}}
+\frac{1}{4}\sqrt{\rho^4-a_3^4}\left(\omega^1\omega^1+\omega^2
  \omega^2+
\frac{|\rho^4-a_1^4|}{\rho^4-a_3^4}\omega^3\omega^3\right)
\label{eq:65a-2}
\end{equation}
for $\rho^2>\max(a_1,a_3)$. This solution corresponds to an axially
symmetric virtual wormhole.

For $\tilde{\phi}_{\text{ T }}$ eq. (2.56) becomes
\begin{eqnarray}
 \frac{d\alpha}{d\lambda}&=&-2\tilde{\phi}_{\text{ T }},\,
  \frac{d\lambda}{dt} = \sqrt{\frac{3\pi}{2}}e^{-3\alpha}\nonumber\\
 \frac{d\beta_+}{d\alpha} &=& -\frac{1}{2}
   \frac{\partial}{\partial\beta_+}\ln|e^{-4\beta_+}-4e^{-\beta_+}|
\label{eq:66a-2}
\end{eqnarray}
$\tilde{\phi}_{\text{ T }}$ and the argument of the log vanish for
$\beta_+=-\frac{2\ln 2}{3}$. It follows from the last of (2.66) that
an initial point at $\alpha=-\infty$, $\beta_+=0$ may either tunnel
to positive $\beta_+$ with increasing $\alpha$ to approach a finite
positive limiting value of $g=e^{2\alpha-4\beta_+}$ while $\alpha$
keeps increasing, or it may tunnel to negative values of $\beta_+$
with $\alpha$ increasing until $\beta_+=-\frac{2\ln 2}{3}$ and
decreasing afterwards, if $\beta_+$ tunnels to values below
$-\frac{2\ln 2}{3}$.  Thus $\tilde{\phi}_{\text{ T }}$ is the solution
of cosmological relevance corresponding to the Hartle Hawking state.
It appears that the NUT region cannot be reached in this state.

\subsection{Inclusion of a cosmological constant}
\label{subsec:2f}
A cosmological term with a cosmological constant $\wedge$ is may be
included in the supersymmetric models, as was shown in \cite{R}. The
Hamiltonian acquires the additional term
\begin{equation}
   H_{\cos} = \wedge
\label{eq:48-2}
\end{equation}
where $\wedge$ is the cosmological constant. We assume that the
conformal factor in $G_{\mu\nu}$ has been chosen in such a way that
(\ref{eq:48-2}) is just constant \cite{R,S}. The supercharges
(\ref{eq:19-2}) get additional contributions
\begin{eqnarray}
   Q_{\cos} &=& i\sqrt{2|\wedge|}\psi^\wedge\nonumber\\
\tilde{Q}_{\cos} &=& \mp i\sqrt{2|\wedge|}\bar{\psi}_\wedge
\label{eq:49-2}
\end{eqnarray}
where the upper (lower) sign applies to positive (negative) $\wedge$,
and where $\psi^\wedge$ and its adjoint $\bar{\psi}_\wedge$
anti-commute with the other $\psi^a$, $\bar{\psi}_a$ and satisfy
\begin{equation}
  \begin{array}{l} (\psi^\wedge)^2=0=(\bar{\psi}_\wedge)^2\\~
    [\psi^{\wedge},\bar{\psi}_\wedge]_+=1 \end{array}
\label{eq:50-2}
\end{equation}
There is still a conserved fermion number $F$ which is now given by
\begin{equation}
   F=\bar{\psi}_a\psi^a+\bar{\psi}_\wedge\psi^\wedge
\label{eq:51-2}
\end{equation}
A state in the sector with $F=f$ can be written as
\begin{equation}
  \begin{array}{l}
   |\psi_f\rangle=\frac{1}{f!}f^{(0)}_{\mu_1\dots\mu_f}
     (q)\bar{\psi}^{\mu_1}\dots\bar{\psi}^{\mu_f}|0\rangle\\
\hspace{2.5cm}+\frac{1}{(f-1)!}f^{(1)}_{\mu_1\dots\mu_{f-1}}(q)
\bar{\psi}_\wedge\bar{\psi}^{\mu_1}\dots\bar{\psi}^{\mu_{f-1}}
   |0\rangle
\end{array}
\label{eq:52-2}
\end{equation}
The constraint $Q|\psi_f\rangle=0$ is expressed by the equations
\begin{equation}
\begin{array}{l}
   \left(\hbar\nabla_\mu-\frac{\partial\phi}{\partial q^\mu}\right)
      {f^{(1)\mu}}_{\mu_1\dots\mu_{f-2}}=0\\
\left(\hbar\nabla_\mu-\frac{\partial\phi}{\partial q^\mu}\right)
    {f^{(0)\mu}}_{\mu_1\dots\mu_{f-1}}+\sqrt{2|\wedge|}
      {f^{(1)}}_{\mu_1\dots\mu_{f-1}}=0\\
\end{array}
\label{eq:53-3}
\end{equation}
and $\tilde{Q}|\psi_f\rangle=0$ takes the explicit form
\begin{equation}
   \epsilon^{\mu_1\dots\mu_{f+1}} \left(\hbar
   \frac{\partial}{\partial q^{\mu_1}}+
     \frac{\partial\phi}{\partial q^{\mu_1}}\right)
      {f^{(1)}}_{\mu_2\dots\mu_f}\mp
       \sqrt{2|\wedge|}{f^{(0)}}_{\mu_1\dots\mu_f}=0
\label{eq:54-2}
\end{equation}
where the doubled sign refers to the choice in eq. (\ref{eq:49-2}).

\subsection{A new conserved probability current?}
\label{subsec:2g}
One hope connected with supersymmetric quantization is to repeat the
success of Dirac's construction of a conserved probability current
which became possible after taking the `square root'of the
Klein-Gordon equation. This hope may be idle; in any case it has not
yet been fulfilled. In order to shed some light on the problems one
encounters we shall discuss here some attempts at the construction of
a conserved current with a positive density. For simplicity we shall
consider the case $G_{\mu\nu}=G_{\mu\nu}^{(0)}=\eta_{\mu\nu}$ and
$\phi$ real. Furthermore we shall put $\hbar=1$. It will be useful to
consider, together with the state vector $|\psi\rangle$, also its
adjoint with respect to the fermionic variables, (but not the
$q$-variables), which we denote by $\langle\psi|$. The scalar product
$\langle\psi|\psi\rangle$ then involves only a summation over the
discrete fermionic components, not an integration over the variables
$q$, i.e.  by construction $\langle\psi|\psi\rangle$ is positive and
$q$-dependent. The supersymmetry constraints are written as
\begin{eqnarray}
   \phantom{i}(Q+\tilde{Q})|\psi\rangle &=& 0\nonumber\\
i(Q-\tilde{Q}) |\psi\rangle &=& 0
\label{eq:55-2}
\end{eqnarray}
We introduce fermionic operator $\xi^\nu$, $\chi^\nu$ by
\begin{eqnarray}
   \xi^\nu &=& \psi^\nu+\bar{\psi}^\nu\nonumber\\
    \chi^\nu &=& i(\psi^\nu-\bar{\psi}^\nu)
\label{eq:56-2}
\end{eqnarray}
with the properties
\begin{equation}
\begin{array}{l}
  \left(\xi^0\right)^+=-\xi^0\quad,\quad\left(\chi^0\right)=-\chi^0\\
     \left(\xi^i\right)^+=\xi^i\quad,\quad\left(\chi^i\right)=
	-\chi^i\quad i=1,2\\~
      [\xi^\nu,\xi^\mu]_+=2\eta^{\nu\mu}=[\chi^\nu,\chi^\mu]_+\\~
	 [\xi^\nu,\chi^\mu]_+=0
\end{array}
\label{eq:57-2}
\end{equation}
Multiplying eqs. (2.74) by $\xi^0$ and $\chi^0$ from the left,
respectively, we obtain
\begin{eqnarray}
\left(-\partial_0+\xi^0\xi^j\partial_j+i\xi^0\chi^\nu
    \phi_{|\nu}\right)|\psi\rangle &=& 0\nonumber\\
\left(-\partial_0+\chi^0\chi^j\partial_j-i\chi^0\xi^\nu
     \phi_{|\nu}\right)|\psi\rangle &=& 0
\label{eq:58-2}
\end{eqnarray}
and the adjoint equations
\begin{eqnarray}
-\partial_0\langle\psi|-\partial_j\langle\psi|\xi^j\xi^0+
  i\phi_{|\nu}\langle\psi|(\chi^\nu)^+\xi^0 &=& 0\nonumber\\
-\partial_0\langle\psi|-\partial_j\langle\psi|\chi^j\chi^0-
  i\phi_{|\nu}\langle\psi|(\xi^\nu)^+\chi^0 &=& 0
\label{eq:59-2}
\end{eqnarray}
Here $\partial_0$, $\partial_j$ and $\phi_{|0}$, $\phi_{|\nu}$ denote
derivatives. Multiplying the first of eqs. (2.77) with
$\langle\psi|A$ from the left and the first of eqs. (2.78) with
$A|\psi\rangle$ from the right, where $A$ is any operator which is
independent of the coordinates, and adding both equations we obtain
\begin{equation}
\begin{array}{l}
-\partial_0\langle\psi|A|\psi\rangle+\langle\psi
  |A\xi^0\xi^j\partial_j|\psi\rangle\\
+ (\partial_j\langle\psi|)\xi^0\xi^jA|\psi\rangle+
   i\phi_{|0}\langle\psi|[A,\xi^0\chi^0]_+|\psi\rangle\\
+i\phi_{|j}\langle\psi|[A,\xi^0\chi^j]_-|\psi\rangle=0
\end{array}
\label{eq:60-2}
\end{equation}
In order to obtain a conservation law the operator $A$ must satisfy
the conditions
\begin{equation}
   [A,\xi^0\xi^j]_-=0=[A,\xi^0\chi^j]_-=[A,\xi^0\chi^0]_+
\label{eq:61-2}
\end{equation}
E.g. the choices $A=i\chi^0$, or $A=i\xi^0\xi^1\xi^2\chi^1\chi^2$
satisfy these conditions.

The conservation law following from eq. (2.79) then reads
\begin{equation}
   \partial_0\langle\psi|A|\psi\rangle+\partial_j
\left(-\langle\psi|A\xi^0\xi^j|\psi\rangle\right)=0.
\label{eq:62-2}
\end{equation}
Similarly, a second class of conservation laws can be derived from
the second of eqs. (2.77)-(2.78). It reads
\begin{equation}
  \partial_0\langle\psi|B|\psi\rangle+\partial_j
   \left(-\langle\psi|B\chi^0\chi^j|\psi\rangle\right)=0
\label{eq:63-2}
\end{equation}
with an operator $B$ independent of the coordinates satisfying
\begin{equation}
   [B,\chi^0\chi^j]_-=[B,\chi^0\xi^j]_-=0=[B,\chi^0\xi^0]_+.
\label{eq:64-2}
\end{equation}
Possible choices are here $B=i\xi^0$ or
$B=i\chi^0\chi^1\chi^2\xi^1\xi^2$.  Unfortunately, among all these
conserved currents there is none with a positive density
$\langle\psi|A|\psi\rangle$ or $\langle\psi|B|\psi\rangle$ because
eqs. (2.80), (2.82) imply
\begin{equation}
  \text{ Tr } A = -\text{ Tr }(\chi^0\xi^0A\xi^0\chi^0)=
   -\text{ Tr } A
\label{eq:65-2}
\end{equation}
and the same equation for $B$ i.e. the operators $A$ and $B$ or any
linear conbination cannot be positive.

Chosing $A=B=1$ we obtain the balance equations
\begin{equation}
\partial_0\langle\psi|\psi\rangle+\partial_j
 (-\langle\psi|\xi^0\xi^j|\psi\rangle)=
  i\phi_{|0}\langle\psi|\xi^0\chi^0|\psi\rangle
\label{eq:66-2}
\end{equation}
and
\begin{equation}
\partial_0\langle\psi|\psi\rangle+\partial_j
  (-\langle\psi|\chi^0\chi^j|\psi\rangle)=
   i\phi_{|0}\langle\psi|\xi^0\chi^0|\psi\rangle.
\label{eq:67-2}
\end{equation}
The $\alpha$-dependence of $\phi$ is seen to spoil the conservation
laws with a positive probability density. Thus, if $\phi=0$ as in
Bianchi type I models, one has conserved currents with a positive
density. On the other hand, subtracting eqs. (2.85), (2.86) we find
the general transversality condition
\begin{equation}
   \partial_j\langle\psi|(\xi^0\xi^j-\chi^0\chi^j)|\psi\rangle=0
\label{eq:68-2}
\end{equation}

\subsection{Interpretation of the wave function}
\label{subsec:2h}
In order to make contact between the equations of motion and physical
reality it is necessary to fix a physical interpretation of the
wave-function. Unfortunatley, there is as yet no generally accepted
interpretation like the usual statistical intepretation of the common
quantum mechanics. Instead several alternatives have been proposed in
the literature (see e.g.[1-7,41-43]), some of which we now discuss
briefly.

\noindent
 {\bf a) Semiclassical interpretation:}

As always in quantum mechanics the wave function can only be
intepreted with respect to a given set-up for a measurement whose
role is, among others, to lift the measured phenomenon from the
microscopic quantum level to a macroscopic level which can be treated
classically. In the case where the wave-function refers to the entire
(microscopic) universe there is no {\bf separate} measurement device,
and the only possibility is apparently that the universe itself acts
as its own measurement device. In this view, therefore, a direct
physical meaning can be given to the wave-function of the universe
only after it has reached a form in which the classical features of
the universe have become apparent. By contrast, the role of the
wave-function in the genuine quantum domain is reduced to a mere
mathematical device allowing to calculate the interpretable
semi-classical wave-function. This is perhaps the most conservative
attitude towards the problem of interpretation. It is contained, in
the semi-classical limit, in all alternative and stronger proposals
trying to give some meaning to the wave-function even in the quantum
domain. The desire to do this arises from the fact that not all
solutions of the Wheeler DeWitt equation reach the semi-classical
regime.

\noindent
{\bf b) Probability current from the time-independent Wheeler DeWitt
equation}:

The (usual time-independent) Wheeler DeWitt equation based on the
(non-supersymmetric) quantization of $H$, eq. (\ref{eq:6-1}), allows
to define a conserved current, like in the Klein-Gordon equation,
with generally non-positive density \cite{B,C}. It can be argued
\cite{C} that sufficiently close to the semi-classical limit the
conserved density is positive to a very good approximation and can
therefore be made the basis of a statistical interpretation. This
interpretation has often been used (e.g. \cite{B,C,G,u,v}). In the
supersymmetric framework of the present paper there are also
fermionic degrees of freedom on which the wave-function depends. The
definition of the conserved current of the Wheeler DeWitt equation
must then be extended and reads
\begin{equation}
   j^\nu=\frac{i}{2}G^{\nu\mu}(q)\left\{\langle\psi|
\frac{\partial}{\partial q^\mu}|\psi\rangle-\left(
\frac{\partial}{\partial q^\mu}\langle\psi|\right)\psi\rangle\right\}
\label{eq:69-2}
\end{equation}
with the scalar product $\langle\psi|\psi\rangle$ taken with respect
to the fermionic variables only, as defined in section \ref{subsec:2g}.
According to Misner \cite{C} the direction (with the unit vector
$n_\nu$) of the flow of time in mini-superspace should be defined by
the condition $n_\nu j^\nu>0$ which can be satisfied, close to the
semi-classical limit, at least for states with a well-defined
classical limit (i.e.  states not containing superpositions of
macroscopically different quantum states).

\noindent
{\bf c) Probability current from the time-dependent Wheeler-DeWitt
equation}:

It is possible to introduce a cosmological time $T$ into the
time-independent Wheeler-DeWitt equation with a cosmologial term. To
this end one chooses the conformal gauge in which the cosmological
term is constant (cf. section \ref{subsec:2f}) and then interprets
this term as a constant of integration, i.e. an energy eigenvalue in
a time-independent Schroedinger equation. It is then natural to look
at the associated time-dependent equation,
\begin{equation}
   i\hbar\frac{\partial|\psi\rangle}{\partial T}=H|\psi\rangle
\label{eq:70-2}
\end{equation}
which is the time-dependent Wheeler-DeWitt equation. It has been
shown that $T$ is given by the elapsed space-time volume (see
\cite{o} and references given there). Eq. (\ref{eq:70-2}) has a
positive conserved invariant measure \cite{o}
\begin{equation}
  dP=\langle\psi|\psi\rangle\sqrt{|\det(G_{\mu\nu})|}d^nq
\label{71-2}
\end{equation}
with the scalar product $\langle\psi|\psi\rangle$ again taken with
respect to the fermionic variables only as defined in section
\ref{subsec:2g}.  $dP$ is therefore a natural choice for a
probability measure (see e.g. \cite{q}). It applies also to solutions
of the time-independent Wheeler DeWitt equation to which we shall
confine ourselves in the present paper. This statistical
interpretation is also frequently used as e.g. in \cite{F,p,q,s,t,u}.
A discussion of conditional probabilities and entropy in quantum
cosmology based on this interpretation has been given in \cite{q}. In
view of the possibility to define a consistent unitary `single
particle' theory of wave-equations of Klein-Gordon type based on this
type of interpretation \cite{p} it appears to be the most convincing
one, at present.

\section{Supersymmetric coupling to a scalar field}\label{sec:3}
\subsection{Expressions for the supercharges in one conformal gauge}
\label{subsec:3a}

Matter can be introduced into the models we are considering by adding
to the Hamiltonian (\ref{eq:6-1}) a matter term $H_M$. For the case
of a spatially homogeneous complex field $z(t)$ with a conventional
form of the kinetic energy $T_M = |\dot{z}|^2$ the form of the
Hamiltonian $H_M$ is fixed by supersymmetry up to an arbitrary
potential $W(z)$ which is an analytic function of $z$. $H_M$ takes
the form \cite{r}
\begin{eqnarray}
  H_M &=& |p_z|^2+V_M(z,z^*)\nonumber\\
  V_M &=& e^{6\alpha+|z|^2}\left(|DW(z)|^2-3|W(z)|^2\right).
\label{eq:1-3}
\end{eqnarray}
Here $p_z=\dot{z}^*$ is the canonically conjugate momentum of $z$,
and $DW(z)=\frac{dW}{dz}+z^*W$.

In the following we shall choose that conformal gauge of the
mini-superspace metric in which the prefactor $\exp (6\alpha+|z|^2)$
in the potential term of eq. (\ref{eq:1-3}) is cancelled. Thus
\begin{equation}
   \begin{array}{l} H_0=\frac{1}{2}G^{\mu\nu}(q)
    \left(p_\nu p_\mu +\frac{\partial\phi}{\partial q^\nu}
     \frac{\partial\phi}{\partial q^\mu}\right)
      +e^{-6\alpha-|z|^2}|p_z|^2\\
\hspace{3.5cm}+\left(|DW(z)|^2-3|W(z)|^2\right)\end{array}
\label{2-3}
\end{equation}
with $G_{\mu\nu}(q)=e^{6\alpha+|z|^2}G_{\mu\nu}^{(0)}$. In this case
the supercharges $Q, \tilde{Q}$ may be extended by matter terms
preserving their property $Q^2=0=\tilde{Q}^2$, so that again a
supersymmetric quantization (\ref{eq:17-2}) is obtained for the
matter-extended Hamiltonian. The construction of $Q$, $\tilde{Q}$ has
been given in ref.
\cite{P} in the context of the Bianchi type IX. However, as the
explicit form of $\phi$ never had to be used there,  the same
construction immediately carries over to the general case. Thus, we
can simply state the result. To be explicit we shall take the case
with configuration space $(\alpha,
\beta_+,\beta_-, z)$. The supercharges may be written as
\begin{eqnarray}
  Q &=& Q_0+Q_K +Q_P\nonumber\\
 \tilde{Q} &=&\tilde{Q}_0+\tilde{Q}_K+\tilde{Q}_P.
\label{eq:3-3}
\end{eqnarray}
Here $Q_0, \tilde{Q}_0$ is the pure gravitational term (without
cosmological constant) considered up to here, and reads explicitely
\begin{eqnarray}
   Q_0 =&&  i e^{-3\alpha-\frac{|z|}{2}^2}
    \Bigg\{\psi^0\left(-\hbar\frac{\partial}{\partial\alpha}+
    \frac{\partial\phi}{\partial\alpha}\right)
    +\psi^1\left(-\hbar\frac{\partial}{\partial\beta_+}
    -3\hbar\bar{\psi}_1\psi^0
    +\frac{\partial\phi}{\partial\beta_+}\right)\nonumber\\
&& +\psi^2\left(-\hbar\frac{\partial}{\partial\beta_-}
    -3\hbar\bar{\psi}_2\psi^0 +\frac{\partial\phi}{\partial\beta_-}
    \right)\Bigg\}\nonumber\\
\tilde{Q}_0 =&&  i e^{-3\alpha-\frac{|z|}{2}^2}
    \Bigg\{-\bar{\psi}_0\left(-\hbar\frac{\partial}{\partial\alpha}-
    \frac{\partial\phi}{\partial\alpha}\right)
    +\bar{\psi}_1\left(-\hbar\frac{\partial}{\partial\beta_+}
    -3\hbar\bar{\psi}_0\psi^1-\frac{\partial\phi}{\partial\beta_+}
    \right)\nonumber\\
&& +\bar{\psi}_2\left(-\hbar\frac{\partial}{\partial\beta_-}
    -3\hbar\bar{\psi}_0\psi^2 -\frac{\partial\phi}{\partial\beta_-}
    \right)\Bigg\}.
\label{eq:4-3}
\end{eqnarray}
$Q_K, \tilde{Q}_K$ in (\ref{eq:3-3}) are kinetic terms due to the
scalar field.  $Q_K$ contains a first new fermionic field $\chi^1$
associated with $z$, and its adjoint $\bar{\chi}_1$ with fermionic
commutation relations, the derivative operators $\partial/\partial
z$, and spin-connection terms following from the extension of
mini-superspace by $z$ and the fact that in our conformal gauge
$G_{\mu\nu}$ depends also on $z$ (while $z^*$ merely plays the role
of a parameter in $Q_K$). Thus
\begin{eqnarray}
 Q_K &=& ie^{-3\alpha-\frac{|z|^2}{2}}\chi^1
  \left(-\sqrt{2}\hbar\frac{\partial}{\partial z}-3\hbar
   \bar{\chi}_1\psi^0-\hbar\frac{z^*}{\sqrt{2}}\psi^a\bar{\psi}_a
    \right) \nonumber\\
\tilde{Q}_K &=& ie^{-3\alpha-\frac{|z|^2}{2}}\bar{\chi}_1
   \left(-\sqrt{2}\hbar\frac{\partial}{\partial z^*}-3\hbar
    \bar{\psi}_0\chi^1-\hbar\frac{z}{\sqrt{2}}\bar{\psi}_a\psi^a
     \right).
\label{eq:5-3}
\end{eqnarray}
Finally, $Q_P$ and $\tilde{Q}_P$ in (\ref{eq:3-3}) are potential
terms due to the scalar field. $Q_P$ contains a second new fermionic
field $\chi^2$ associated with $z$ and $\bar{Q}_P$ contains its
adjoint $\bar{\chi}_2$.  In addition a new second fermionic variable
$\chi^0$ associated with $q^0=\alpha$ also appears in $Q_P$, together
with its adjoint $\bar{\chi}_0$. It is made necessary by the negative
term in the matter potential of (\ref{eq:1-3}). Explicitely
\begin{eqnarray}
   Q_P &=& i\sqrt{2}\chi^2\left\{(DW(z))^*+\frac{\hbar}{\sqrt{3}}
      e^{-3\alpha-\frac{|z|^2}{2}}\bar{\chi}_0\chi^1\right\}
      +i\sqrt{6}\chi^0(W(z))^*\nonumber\\
\tilde{Q}_P &=& i\sqrt{2}\bar{\chi}_2\left(-DW(z)+\frac{\hbar}
      {\sqrt{3}}e^{-3\alpha-\frac{|z|^2}{2}}\bar{\chi}_1\chi^0\right)
      +i\sqrt{6}\bar{\chi}_0W(z)
\label{eq:6-3}
\end{eqnarray}
There is still a conserved fermion number
\begin{equation}
   F=\bar{\psi}_a\psi^a+\bar{\chi}^0\chi_0+\bar{\chi}_1\chi^1+
   \bar{\chi}_2\chi^2
\label{eq:7-3}
\end{equation}
but now there are seven sectors $F=f$ with $f= 0, 1, 2, 3, 4, 5, 6$.

Nontrivial solutions in the sectors $f=0$ and $f=6$ exist only if
$W(z)\equiv 0$ and are then given by
\begin{eqnarray}
 |\psi_0\rangle &=& f(z)e^{-\phi/\hbar}|0\rangle\nonumber\\
 |\psi_6\rangle &=& g(z^*)e^{\phi/\hbar}|n\rangle
\label{eq:8-3}
\end{eqnarray}
with arbitrary analytical functions $f(z)$, $g(z^*)$.

The other fermion sectors can be considered like in section
\ref{subsec:2d}. Thus it is sufficient to consider $f=1, 2, 3$.

In the appendix we show that all solutions in the 1-fermion sector
are of the form (\ref{eq:46-2}). The ansatz $|\psi_1\rangle=\tilde{Q}
f(q,z,z^*)e^{-\phi/\hbar}|0\rangle$ satisfies
$\tilde{Q}|\psi_1\rangle=0$, and the remaining condition
$Q|\psi_1\rangle=0$ leads to the Wheeler-DeWitt equation for
$f(q,z,z^*)$
\begin{eqnarray}
&& \frac{\hbar}{2}\eta^{\mu\nu}
   \left(\hbar\frac{\partial}{\partial q^\mu}-2
   \frac{\partial\phi}{\partial q^\mu}\right)
   \frac{\partial f}{\partial q^\nu}-3\hbar^2
   \frac{\partial f}{\partial\alpha}+\hbar^2
   \frac{\partial^2 f}{\partial z\partial z^*}+\hbar^2 z^*
   \frac{\partial f}{\partial z^*} \nonumber\\
&& \hspace{4cm} = e^{6\alpha+|z|^2}(|DW| ^2-3|W|^2) f.
\label{eq:9-3}
\end{eqnarray}
For $f=2$ the ansatz
\begin{equation}
|\Psi_2\rangle= \tilde{Q}
\left(f^\mu\bar{\psi}_\mu+g_1\bar{\chi}_1+g_2\bar{\chi}_2\right)
      |0\rangle.
\label{eq:10-3}
\end{equation}
can be made, with five undetermined functions $f^\mu, g_1, g_2$.
Terms with $\tilde{Q}\bar{\chi}_0$ have been eliminated from
(\ref{eq:10-3}) by subtracting the vanishing state $0=\tilde{Q}^2
h(q)|0\rangle$ with appropriately chosen $h(q)$. We obtain a system of
four Wheeler-DeWitt equations, i.e.  second order wave-equations, and
two auxiliary equations which are of first order. One of the
auxiliary equations is not independent, however, and may be dropped.
These equations look tedious and we shall not record them here.

Similarly for $f=3$ the ansatz
\begin{equation}
|\Psi_3\rangle = \tilde{Q}\left(f^{\mu\nu}\bar{\psi}_\mu
  \bar{\psi}_\nu+g_1^\mu\bar{\psi}_\mu\bar{\chi}_1+ g_2^\mu
  \bar{\psi}_\mu\bar{\chi}_2+g_{12}\bar{\chi}_1\bar{\chi}_2\right)
  |0\rangle
\label{eq:11-3}
\end{equation}
can be made with 10 undetermined functions $f^{\mu\nu}=-f^{\nu\mu},
g_1^\mu, g_2^\mu, g_{12}$. Again we made use of $\tilde{Q}^2=0$ to
eliminate the five terms involving $\tilde{Q}\bar{\chi}_0$. Now a
system of 6 second-order wave-equations is obtained together with 9
auxiliary equations of first order, five of which are not independent
and may be dropped. The wave-equation (\ref{eq:9-3}) is the simplest
one obtained in any of the nontrivial fermion sectors. To avoid the
singularity for $\phi\rightarrow\infty$ we demand that $\partial
f/\partial q^\nu\rightarrow 0$ in the same limit. The semi-classical
limit $\hbar\rightarrow 0$ coincides with the limit
$e^{6\alpha+|z|^2}\rightarrow\infty$ in dimensionless units. In the
semi-classical limit one may write
\begin{equation}
   fe^{-\phi}\sim e^{iS(q,z,z^*)/\hbar}
\label{eq:12-3}
\end{equation}
to obtain
\begin{equation}
   \frac{1}{2}\eta^{\mu\nu}\frac{\partial S}{\partial q^\mu}
   \frac{\partial S}{\partial q^\nu}+\frac{\partial S}{\partial z}
   \frac{\partial S}{\partial z^*} + V(\alpha,\beta_+,\beta_-)
   +V_M(z,z^*)=0
\label{eq:13-3}
\end{equation}
which is the Hamilton-Jacobi equation of the classical model. Thus
for $\alpha\rightarrow\infty$ a wave-packet moving along trajectories
of the classical system is obtained.

The classical trajectories will start with `initial' conditions in
that region of configuration space where the wave-function makes its
transition from an exponentially decaying or growing behavior $\sim
e^{-\phi/\hbar}$ or $e^{-\tilde{\phi}/\hbar}$to an oscillatory
behavior $(\sim e^{iS/\hbar})$ \cite{F,s,G}. E.g.  if a Bianchi type
IX space is considered (without invoking supersymmetry this was done
in \cite{t,u,v,w}), and if the potential $W(z)$ is chosen in such a
way that the transition from exponential to oscillatory behavior
occurs for sufficiently large $e^{2\alpha}>>1$, then the
corresponding classical `initial' values of $\beta_+,\beta_-$ and
$\partial\beta_+/\partial\alpha$, $\partial\beta_-/\partial\alpha$
will be very small, providing a possible explanation for the observed
isotropy of the universe.

\subsection{Choice of flat mini-superspace metric}
\label{subsec:3b}
The choice of a prefactor in $G_{\mu\nu}$ is just a matter of
convenience \cite{C}. Therefore it is of interest to explore other
choices than the one  made in the preceeding section. The simplest
and most obvious one seems to be the choice
$G_{\mu\nu}(q)=G_{\mu\nu}^{(0)}=\eta_{\mu\nu}$. In the present
section we consider the coupling to a scalar field for this case and
derive expressions for $Q$, $\tilde{Q}$ corresponding to eqs.
(\ref{eq:3-3}-\ref{eq:6-3}). The Hamiltonian $H_0$ now takes the form
\begin{equation}
   H_0=\frac{1}{2}\eta^{\mu\nu} \left(p_\mu
   p_\nu+\frac{\partial\phi}{\partial q^\mu}
   \frac{\partial\phi}{\partial q^\nu}\right) + H_M
\label{eq:14-3}
\end{equation}
where $H_M$ is given by eq. (\ref{eq:1-3}). The supercharges are
again written as in eq. (\ref{eq:3-3}). The expressions for $Q_0$,
$\tilde{Q}_0$ are now simpler than in eq. (\ref{eq:4-3}) because no
connection terms appear for a flat metric. Thus
\begin{equation}
 \begin{array}{l} Q_0=i\Bigg\{
  \psi^0\left(-\hbar\frac{\partial}{\partial\alpha}+
   \frac{\partial\phi}{\partial\alpha}
    \right)
+\psi^1\left(-\hbar\frac{\partial}{\partial\beta_+}+
   \frac{\partial\phi}{\partial\beta_+}
    \right)
+\psi^2\left(-\hbar\frac{\partial}{\partial\beta_-}+
  \frac{\partial\phi}{\partial\beta_-}
   \right) \Bigg\}\\
\tilde{Q}_0=i\Bigg\{
   \bar{\psi}_0\left(\hbar\frac{\partial}{\partial\alpha}+
    \frac{\partial\phi}{\partial\alpha}
     \right)
+\bar{\psi}_1\left(-\hbar\frac{\partial}{\partial\beta_+}-
     \frac{\partial\phi}{\partial\beta_+}
      \right)
+\bar{\psi}^2\left(-\hbar\frac{\partial}{\partial\beta_+}-
      \frac{\partial\phi}{\partial\beta_-} \right) \Bigg\}.
\end{array}
\label{eq:15-3}
\end{equation}
Also the kinetic matter terms $Q_K$, $\tilde{Q}_K$ now lack all
3-fermion terms, because the extension of mini-superspace by $z$ and
$z^*$ leaves the extended metric flat. Thus
\begin{eqnarray}
   Q_K &=& i\sqrt{2}\chi^1\left(-\hbar
   \frac{\partial}{\partial z}\right) \nonumber\\
\tilde{Q}_K &=& i\sqrt{2}\bar{\chi}_1\left(-\hbar
    \frac{\partial}{\partial z^*}
     \right).
\label{eq:16-3}
\end{eqnarray}
However, the potential terms $Q_P$, $\tilde{Q}_P$ now become more
complicated. The additional dependence of $H_M$ on $\alpha$ and
$|z|^2$ entails the necessity of more 3-fermion terms which do not
have any obvious interpretation in terms of spin-connections. Thus we
write
\begin{equation}
\begin{array}{l}
Q_P = i\sqrt{2}e^{3\alpha+|z|^2/2}
\left(\chi^2(DW(z))^*+\sqrt{3}\chi^0(W(z))^*\right)+i\hbar T\\
\tilde{Q}_P = i\sqrt{2}e^{3\alpha+|z|^2/2}
       \left(-\bar{\chi}_2DW(z))+\sqrt{3}\bar{\chi}_0(W(z))\right)
-i\hbar\tilde{T}
\end{array}
\label{eq:17-3}
\end{equation}
where $T$ and $\tilde{T}$ denote 3-fermion terms. They can be found
by compensating all terms in $Q^2=0=\tilde{Q}^2$.The result is
\begin{equation}
\begin{array}{l}
   T = \sqrt{\frac{2}{3}}\bar{\chi}_0\chi^1\chi^2+
     3\bar{\chi}_2\psi^0\chi^2+3\bar{\chi}_0\psi^0\chi^0
      +\frac{z^*}{\sqrt{2}}\left(\bar{\chi}_0\chi^1\chi^0+
      \bar{\chi}_2\chi^1\chi^2\right)\\
\tilde{T}= -\sqrt{\frac{2}{3}}\bar{\chi}_2\bar{\chi}_1\chi^0-
     3\bar{\chi}_2\bar{\psi}_0\chi^2-3\bar{\chi}_0\bar{\psi}_0\chi^0
     +\frac{z}{\sqrt{2}}\left(\bar{\chi}_0\bar{\chi}_1\chi^0+
     \bar{\chi}_2\bar{\chi}^1\chi^2\right)
\end{array}
\label{eq:18-3}
\end{equation}
In the zero- and 1-fermion sector the 3-fermion terms do not
contribute. Proceeding as in eqs. (\ref{eq:8-3}-\ref{eq:9-3}) the
wave-equation in the 1-fermion sector is obtained as
\begin{equation}
\begin{array}{l}
  \frac{1}{2}\hbar\eta^{\mu\nu}
   \left(\hbar\frac{\partial}{\partial q^\mu}-2
    \frac{\partial\phi}{\partial q^\mu}\right)
     \frac{\partial f}{\partial q^\nu}+\hbar^2
      \frac{\partial^2f}{\partial z\partial z^*}=
       e^{6\alpha+|z|^2}(|DW|^2-3|W|^2)f.  \end{array}
\label{eq:19-3}
\end{equation}
It is somewhat simpler in form than eq. (\ref{eq:9-3}) but leads to
the same conclusions. In particular, the semi-classical solutions
(\ref{eq:12-3}), (\ref{eq:13-3}) are the same.

\subsection{Approximate solutions in the 1-fermion sector}
\label{subsec:3c}
For $\alpha$ negative and sufficiently large the right-hand side of
eq. (\ref{eq:19-3}) is negligible and also $\phi$ approaches zero.
Then
\begin{equation}
   f\simeq e^{ik_\nu q^\nu+\frac{i}{\sqrt{2}}(kz+k^*z^*)}
\label{eq:25-3}
\end{equation}
with
\begin{equation}
  k_0^2=|k|^2+k_1^2+k_2^2
\label{eq:26-3}
\end{equation}
 and
\begin{equation}
  |\psi_1\rangle\simeq e^{-\phi}e^{ik_\nu q^\nu+
   \frac{i}{\sqrt{2}}(kz+k^*z^*)} (\hbar k_\nu\bar{\psi}^\nu+\hbar
    k^*\bar{\chi}_1) |0\rangle
\label{eq:27-3}
\end{equation}

For sufficiently large $\alpha$ (see below) eq. (\ref{eq:19-3}) may
be solved in a Born-Oppenheimer approximation assuming that the
scalar field adjusts itself quasi-instantaneously to the
gravitational field. We consider the case where $W(z)$ has a
quadratic stationary point $z_0$ with $W(z_0)=0$,
\begin{eqnarray}
  W(z) &=& \frac{c}{2}(z-z_0)^2+\dots\nonumber\\
  DW  &=& c(z-z_0)+\dots,\quad c\neq 0.
\label{eq:28a-3}
\end{eqnarray}
Thus
\begin{equation}
  V_M\simeq e^{6\alpha+|z_0|^2}\left(|c|^2|z-z_0|^2+\dots\right).
\label{eq:28-3}
\end{equation}
For this potential and for fixed $\alpha$ we can solve the eigenvalue
problem
\begin{equation}
-\hbar^2\frac{\partial\tilde{f}}{\partial z\partial z^*}+V_Mf=
  E(\alpha)\tilde{f}
\label{eq:29-3}
\end{equation}
in terms of eigenfunctions and eigenvalues of the two-dimensional
harmonic oscillator. The energy $E$ depends on $\alpha$ like
\begin{equation}
  E(\alpha)=E_0e^{3\alpha}.
\label{eq:30-3}
\end{equation}
The constant prefactor $E_0$ depends on the two quantum numbers of
the harmonic oscillator which will not change in the adiabatic
regime. It then remains to solve the reduced problem
\begin{equation}
  \frac{\hbar}{2}\eta^{\mu\nu}
\left(\hbar\frac{\partial}{\partial q^\mu}-2
 \frac{\partial\phi}{\partial q^\mu}\right)
 \frac{\partial g}{\partial q^\nu}=E_0e^{3\alpha}g
\label{eq:31-3}
\end{equation}
where the right-hand side gives the energy of the matter field. The
assumption of adiabaticity is satisfied if the {\it relative} change
of the frequency $\omega(\alpha)=2|c|e^{3\alpha+|z_0|^2/2}$ of the
scalar field over a period $2\pi/\omega(\alpha)=T$ remains small.
Using $\alpha$ itself as a parameter to measure time we have for the
change $\Delta\alpha$ of $\alpha$ over one period
\begin{eqnarray}
 && \int^{\Delta\alpha}_0\;\omega(\alpha)d\alpha=2\pi\nonumber\\
 && \Delta\alpha\simeq\frac{3\pi}{|c|}e^{-3\alpha-|z_0|^2/2}
\label{eq:32-3}
\end{eqnarray}
and the adiabatic condition becomes
\begin{equation}
  3\Delta\alpha<<1
\label{eq:33-3}
\end{equation}
which will be satisfied in the limit of a large scale-parameter.

The total wave-function in this limit then is approximately
\begin{equation}
f(z,z^*,q)\simeq\tilde{f}(z,z^*|\alpha)g(\alpha,\beta_+,\beta_-)
\label{eq:34-3}
\end{equation}
where $\tilde{f}$ solves eq. (3.25) and $g$ solves
eq.(\ref{eq:31-3}). For concreteness let us assume now that we are
dealing with the Bianchi type IX case. In the limit where the
adiabatic approximation is valid the prefactor $e^{-\phi/\hbar}$ or
$e^{-\tilde{\phi}/\hbar}$ of $f$ will then be very sharply peaked at
$\beta_+=0=\beta_-$. Therefore we may put $\beta_+=\beta_-=0$ in $f$
and eq. (\ref{eq:31-3}) is reduced, respectively, to
\begin{equation}
 -\frac{\hbar^2}{2}\frac{\partial^2 g}{\partial\alpha^2}+
   \hbar e^{2\alpha}\frac{\partial g}{\partial\alpha}=
    E_0e^{3\alpha}g \quad\text{ for }\; \phi=\phi_9
\label{eq:35-3}
\end{equation}
or
\begin{equation}
  -\frac{\hbar^2}{2}\frac{\partial^2 g}{\partial\alpha^2}-
   \hbar e^{2\alpha}\frac{\partial g}{\partial\alpha}=
    E_0e^{3\alpha}g \quad\text{ for }\; \phi=\tilde{\phi}_9
\label{eq:35b-3}
\end{equation}
In the semi-classical limit the solutions are given by
\begin{equation}
  g\simeq(2E_0e^{3\alpha}-e^{4\alpha})^{1/2}\exp
  \left[\frac{1}{2}e^{2\alpha}-i\int^\alpha\;d\alpha
  (2E_0e^{3\alpha}-e^{4\alpha})^{1/2}\right]
\label{eq:36-3}
\end{equation}
for $\phi=\phi_9$ or
\begin{equation}
  g\simeq(2E_0e^{3\alpha}-e^{4\alpha})^{1/2}\exp
  \left[-\frac{1}{2}e^{2\alpha}-i\int^\alpha\;d\alpha
  (2E_0e^{3\alpha}-e^{4\alpha})^{1/2}\right]
\label{eq:36b-3}
\end{equation}
for $\phi=\tilde{\phi}_9$, which are both outgoing waves as long as
\begin{equation}
 e^\alpha<2E_0.
\label{eq:37-3}
\end{equation}
This condition in the present model defines the maximum radius of the
universe, where the outgoing wave is reflected to become a standing
wave. In the physical universe, long before this event happens, the
coherence of the wave-function has been spread over so many degrees
of freedom (not contained in our mini-superspace) that this coherence
becomes completely irrelevant for any conceivable physical process
and a classical description is required.

\vspace{0.25cm}
\noindent
{\Large\bf Acknowledgements}\\ This work was supported by the
Deutsche Forschungsgemeinschaft through the Sonderforschungsbereich
237 ``Unordnung und gro{\ss}e Fluktuationen'', and by the
``Deutsch-\newline Ungarisches Kooperationsabkommen'' through the
grant X231.3. One of the authors (R.~G.) would like to thank Dr.
P.~D.~D'Eath and Dr.~G.~W.~Gibbons for useful discussions. We also
want to thank Dr.~Gibbons for bringing the important work  of refs.
\cite{j,l} to our attention.

\appendix
\section{}\label{sec:ap}
In this appendix we examine the general form of the solution of
$\tilde{Q}|\psi_1\rangle=0$ in the 1-fermion sector with
$\tilde{Q}=\tilde{Q}_0+\tilde{Q}_K+\tilde{Q}_P$ given by
eqs.~(\ref{eq:4-3})-(\ref{eq:6-3}). With
$|\psi_1\rangle=(f^\nu\bar{\psi}_\nu+g^\nu\bar{\chi}_\nu)|0\rangle$
we obtain that the 15 coefficients of $\bar{\psi}_\nu\bar{\psi}_\mu$,
$\bar{\chi}_\nu\bar{\psi}_\mu$ and $\bar{\chi}_\nu\bar{\chi}_\mu$
must vanish. Using the abbreviations
\begin{eqnarray}
 \Omega &=& e^{-3\alpha-\frac{|z|^2}{2}}\nonumber\\
 D_\nu &=& -\hbar\partial_\nu-\partial\phi/\partial q^\nu
\label{eq:1a}
\end{eqnarray}
the conditions that the coefficient of $\bar{\psi}_\nu\bar{\psi}_\mu$
vanish read
\begin{eqnarray}
 && -D_\alpha f^1-D_{\beta_+}f^0+3\hbar f^1=0\nonumber\\
 && -D_\alpha f^2-D_{\beta_-}f^0+3\hbar f^2=0\nonumber\\
 && D_{\beta_+}f^2-D_{\beta_-}f^1=0
\label{eq:2a}
\end{eqnarray}
Defining the functions $\tilde{f}^\nu$ by
\begin{equation}
  f^\nu=e^{-3\alpha-\phi/\hbar}\tilde{f}^\nu
\label{eq:3a}
\end{equation}
eqs. (A2) become
\begin{eqnarray}
 \partial_\alpha\tilde{f}^1+\partial_{\beta_+}\tilde{f}^0 &=& 0
  \nonumber\\
 \partial_\alpha\tilde{f}^2+\partial_{\beta_-}\tilde{f}^0 &=&0
  \nonumber\\
 \partial_{\beta_+}\tilde{f}^2-\partial_{\beta_-}\tilde{f}^1 &=&0
\label{eq:4a}
\end{eqnarray}
which imply
\begin{equation}
  \tilde{f}^0 = -\frac{\partial F}{\partial\alpha},\;
  \tilde{f}^1=\frac{\partial F}{\partial\beta_+},\; \tilde{f}^2 =
   \frac{\partial F}{\partial\beta_-}
\label{eq:5a}
\end{equation}
or
\begin{equation}
  f^\nu = i\Omega D^\nu\tilde{F}
\label{eq:6a}
\end{equation}
with
\begin{equation}
  \tilde{F}=\frac{i}{\hbar}e^{|z|^2/2}e^{-\phi/\hbar}F.
\label{eq:7a}
\end{equation}
Here $F$ and $\tilde{F}$ are yet undetermined functions of
$\alpha,\beta_+,\beta_-,z,z^*$. Eq.~(A6) implies
\begin{equation}
   f^\nu\bar{\psi}_\nu|0\rangle=\tilde{Q}_0\tilde{F}|0\rangle.
\label{eq:8a}
\end{equation}
Next we consider the conditions that the coefficients of
$\bar{\chi}_\nu\bar{\chi}_\mu$ vanish. They take the form
\begin{eqnarray}
  i\sqrt{2}\hbar\Omega\partial_{z^*}g^0+i\sqrt{6}Wg^1&=&0\nonumber\\
  i\sqrt{2}DWg^0+i\sqrt{6}Wg^2&=& 0\nonumber\\
  -i\sqrt{2}\hbar\Omega\partial_{z^*}g^2+i\sqrt{2}DWg^1
  -i\sqrt{\frac{2}{3}}\hbar\Omega g^0 &=& 0.
\label{eq:9a}
\end{eqnarray}
{}From the first two of these equations it follows (for $W\neq 0$) that
\begin{eqnarray}
  g^1 &=& -\frac{\hbar}{\sqrt{3}}\frac{\Omega}{W}\partial_{z^*}g^0
   \nonumber\\
  g^2 &=& -\frac{1}{\sqrt{3}}\frac{DW}{W} g^0
\label{eq:10a}
\end{eqnarray}
and the last of eqs.~(A9) is then automatically satisfied. With
\begin{equation}
   \tilde{g}^0=-\frac{i}{\sqrt{6}}\frac{1}{W}g^0
\label{eq:11a}
\end{equation}
eqs.~(A10) imply
\begin{equation}
  g^\nu\bar{\chi}_\nu|0\rangle=(\tilde{Q}_K+\tilde{Q}_P)\tilde{g}^0
  |0\rangle.
\label{eq:12a}
\end{equation}
Finally we equate to zero the coefficients of
$\bar{\chi}_0\bar{\psi}_\nu$ which yields
\begin{equation}
  i\Omega D^\nu g^0-i\sqrt{6}Wf^\nu=0
\label{eq:13a}
\end{equation}
and implies with eqs.~(A11), (A6) that
\begin{equation}
  D^\nu(\tilde{g}^0-\tilde{F})=0.
\label{eq:14a}
\end{equation}
Hence, we may rewrite eq.~(A8) as
\begin{equation}
  f^\nu\bar{\psi}_\nu|0\rangle=\tilde{Q}_0\tilde{g}^0|0\rangle
\label{eq:15a}
\end{equation}
and eqs.~(A12), (A15) together imply
\begin{equation}
  |\psi_1\rangle=(f^\nu\bar{\psi}_\nu+g^\nu\bar{\chi}_\nu)|0\rangle=
  (\tilde{Q}_0+\tilde{Q}_K+\tilde{Q}_P)\tilde{g}^0|0\rangle=
  \tilde{Q}\tilde{g}^0|0\rangle.
\label{eq:16a}
\end{equation}
Due to the form (A16) of $|\psi_1\rangle$ the coefficients of the
remaining terms $\bar{\chi}_\mu\bar{\psi}_\nu$, $(\mu\neq 0)$ now
automatically vanish. Nontrivial solutions in the 1-fermion sector
therefore must all have the form (A16), which is permitted only for
a non-definite signature of the metric in mini-superspace. For a
Riemannian signature of the metric there are no states in the
1-fermion sector.

\end{document}